\documentclass[journal]{IEEEtran}
\usepackage{amsmath,amsfonts}
\usepackage{amsthm}
\usepackage{algorithmic}
\usepackage{multirow}
\usepackage{booktabs}
\usepackage[ruled,linesnumbered]{algorithm2e}
\usepackage{array}
\usepackage[caption=false,font=normalsize,labelfont=sf,textfont=sf]{subfig}
\usepackage{textcomp}
\usepackage{stfloats}
\usepackage{url}
\usepackage{verbatim}
\usepackage{graphicx}
\usepackage{cite}
\usepackage{cancel}
\usepackage{indentfirst}
\usepackage{epstopdf}
\usepackage{cite}
\usepackage{color}
\usepackage[colorlinks,citecolor=blue,linkcolor=blue,urlcolor=blue]{hyperref}

\begin{document}
	
	\title{	\LARGE  A Hybrid Dynamic Subarray Architecture for Efficient DOA Estimation in THz Ultra-Massive Hybrid MIMO Systems}
	\author{Ye Tian,~Jiaji Ren,~Tuo Wu,~Wei Liu,~Chau Yuen,~\IEEEmembership{Fellow,~IEEE,} M\'{e}rouane Debbah,  \emph{Fellow},  \emph{IEEE},  	\\Naofal Al-Dhahir, \emph{Fellow, IEEE},   Matthew C. Valenti, \emph{Fellow},   \emph{IEEE},  Hing Cheung So, \emph{Fellow, IEEE} \\  and Yonina C. Eldar,  \emph{Fellow},  \emph{IEEE} 
		\vspace{-2mm}  \thanks{ \emph{(Corresponding author: Tuo Wu.)}
			
			Ye Tian and Hongyun Zhao are with the Faculty of Electrical Engineering and Computer Science, Ningbo University, Ningbo 315211, China (E-mail: $\rm  tianfield@126.com; 2311100053@nbu.edu.cn$).
			T. Wu and C. Yuen are with the School of Electrical and Electronic Engineering, Nanyang Technological University, 639798, Singapore (E-mail: $\rm \{tuo.wu, chau.yuen\}@ntu.edu.sg$).
			Wei Liu is with the Department of Electrical and Electronic Engineering, Hong Kong Polytechnic University, Kowloon, Hong Kong, China (E-mail: $\rm wliu.eee@gmail.com$).  
			M. Debbah is with  KU 6G Research Center, Department of Computer and Information Engineering, Khalifa University, Abu Dhabi 127788, UAE (E-mail: $\rm merouane.debbah@ku.ac.ae$).
			Naofal Al-Dhahir is with the Department of Electrical and Computer Engineering, The University of Texas at Dallas, Richardson, TX 75080 USA (E-mail: $\rm aldhahir@utdallas.edu$).
			M. Valenti is with the Lane Department of Computer Science and Electrical Engineering, West Virginia University, Morgantown, USA (E-mail: $\rm valenti@ieee.org$).
			H. C. So is with the Department of Electrical Engineering City University of Hong Kong, Hong Kong, China. (E-mail: $\rm hcso@ee.cityu.edu.hk$).
			Yonina C. Eldar is with Weizmann Institute of Science, Israel (E-mail: $ \rm yonina.eldar@weizmann.ac.il$).

		} 
	}
	\markboth{IEEE TRANSACTIONS ON SIGNAL PROCESSING,~Vol.~XX, No.~XX, XX~XXXX}%
	{Shell \MakeLowercase{\textit{et al.}}: A Sample Article Using IEEEtran.cls for IEEE Journals}
	
	
	\maketitle 
	\begin{abstract}
		Terahertz (THz) communication combined with ultra-massive multiple-input multiple-output (UM-MIMO) technology is promising for 6G wireless systems, where fast and precise direction-of-arrival (DOA) estimation is crucial for effective beamforming. However, finding DOAs in THz UM-MIMO systems faces significant challenges: while reducing hardware
		complexity, the hybrid analog-digital (HAD) architecture introduces inherent difficulties in spatial information acquisition the large-scale antenna array causes significant deviations in eigenvalue decomposition results; and conventional two-dimensional DOA estimation methods incur prohibitively high computational overhead, hindering fast and accurate realization. To address these challenges, we propose a   hybrid dynamic subarray (HDS) architecture that strategically divides antenna elements into subarrays, ensuring phase differences between subarrays correlate exclusively with single-dimensional DOAs. Leveraging this architectural innovation, we develop two efficient algorithms for DOA estimation: a reduced-dimension MUSIC (RD-MUSIC) algorithm that enables fast processing by correcting large-scale array estimation bias, and an improved version that further accelerates estimation by exploiting THz channel sparsity to obtain initial closed-form solutions through specialized two-RF-chain configuration. Furthermore, we develop a   theoretical framework through Cram\'{e}r-Rao lower bound analysis, providing fundamental insights for different HDS configurations. Extensive simulations demonstrate that our solution achieves both superior estimation accuracy and computational efficiency, making it particularly suitable for practical THz UM-MIMO systems.
		
	\end{abstract}
	
	\vspace{-3mm}
	\section{Introduction} 
	\IEEEPARstart{T}{ERAHERTZ} (THz)   technology anticipated to be  an important part  of 6G wireless communications \cite{r1}. While the ultra-wide bandwidth of terahertz bands promises unprecedented data transmission rate, it also introduces severe propagation attenuation that significantly limits the effective communication range \cite{r2}. This fundamental challenge in terahertz communications, however, can be addressed through the deployment of ultra-massive multiple-input multiple-output (UM-MIMO) antenna arrays. The sub-millimeter wavelengths characteristic of the terahertz band enable thousands of miniature antenna elements to be densely integrated within a limited physical space. By leveraging these large-scale arrays and employing beamforming techniques for precise directional transmission, the antenna array gain can effectively compensate for the propagation loss, thereby extending the communication range to useful distances.

	Despite the potential of THz UM-MIMO systems to enhance wireless communication system performance, their practical implementation still faces significant challenges. Specifically, even slight beam misalignment can lead to substantial signal power loss and significant degradation of system performance \cite{r15,r16}. Therefore, achieving high-precision direction-of-arrival (DOA) estimation is a key requirement for ensuring the performance of THz UM-MIMO systems, since it is a critical step in compensating for beam misalignment. Numerous high-accuracy and high-resolution DOA estimation algorithms have been investigated in the fields of wireless communications, radar, sonar, and astronomy. Examples include the estimation of signal parameters via rotational invariance techniques (ESPRIT) \cite{r18,r19}, compressed sensing-based algorithms \cite{r20}, spatial structure matching-based algorithms  \cite{r21}, deep learning-based algorithms \cite{r22,r23,r24}, and general asymptotic theory (GAT)-based algorithms \cite{r25,r26}.  These algorithms are effective   for high-performance DOA estimation in large-scale arrays; however, they are all designed for fully digital (FD)  architectures. The traditional FD connection structure suffers from high hardware costs  \cite{r5} because each antenna   typically requires a dedicated radio frequency (RF) chain, including components such as analog-to-digital converters (ADCs), mixers, and power amplifiers \cite{r3, r4}.
	
	To tackle this issue, hybrid analog/digital (HAD) architectures have been proposed in recent years as an effective solution for reducing the number of RF chains while maintaining system performance. Generally, HAD architectures are classified into three categories: hybrid fully-connected (HFC)   \cite{r6,r7}, hybrid subarray (HS)    \cite{r8,r9,r10}, and hybrid dynamic subarray (HDS)   \cite{r11,r12,r13,r14}. Among them, the HDS architecture stands out for its excellent flexibility in array design, making it particularly suitable for various practical   scenarios. Leveraging the feature of excellent flexibility,  \cite{r32}  proposed a two-dimensional (2-D) multiple signal classification (MUSIC)-based algorithm that achieves millimeter-level DOA estimation accuracy in THz UM-MIMO systems. However, it requires computationally demanding 2-D search,  leaving room for improvement in terms of enhancing flexibility and reducing computational resource consumption.
	
	To simultaneously achieve speed and accuracy when applying HAD in THz UM-MIMO systems, four non-trivial technical issues need to be addressed concurrently.  First, the HAD architecture design requires a balance between hardware complexity and system performance. While reducing RF chains can effectively lower system costs, it introduces challenges in fast spatial information acquisition and precise phase coherence maintenance across subarrays. Consequently, the design of subarray partitioning and switch networks poses a fundamentally complicated problem. Second, achieving accurate DOA estimation under the HAD structure faces significant difficulties in large-scale scenarios. The deviation between eigenvalue decomposition results and true values necessitates intricate correction mechanisms. Moreover, the limited number of RF chains makes rapid spatial information extraction an inherent issue, especially when directly applying traditional subspace-based methods. Third, the high complexity of traditional 2-D spectral search methods makes DOA estimation difficult to process in real-time in THz UM-MIMO systems, and even slight beam misalignment can lead to severe performance degradation. Fourth, establishing rigorous theoretical  analysis is also an arduous task. The complex coupling between different architectures and its impact on estimation speed and accuracy make it difficult to derive analytical bounds and equivalence relationships, thereby hindering theoretical guidance for practical system design and optimization.
	
	Motivated by these aboved issues, this paper presents a novel HDS architecture enabling efficient DOA estimation for THz UM-MIMO systems. By introducing a specially designed flexible HDS structure and developing fast yet precise estimation approach, our solution achieves both high accuracy and low computational complexity. The main contributions of this work are summarized as follows:

	\begin{itemize}
		\item \textbf{\textit{Novel Architecture:}} We devise a  HDS architecture with enhanced flexibility. In this architecture, antenna elements are strategically divided into multiple subarrays based on RF chain count, ensuring phase differences between subarrays correlate exclusively with either azimuth or elevation DOA. This sophisticated connection scheme achieves an optimal trade-off among design flexibility, system overhead, and computational complexity of DOA estimation, addressing the first challenge of balancing hardware complexity and system performance. 
		
		\item \textbf{\textit{Efficient Algorithms:}} To tackle the second and third challenges of precise and fast DOA estimation, we develop two low-complexity yet high-accuracy DOA estimators for the designed architecture. First, we propose a phase compensation based reduced-dimension MUSIC (RD-MUSIC) algorithm that effectively balances computational overhead with estimation accuracy demands in terahertz communications. Second, by exploiting the inherent sparsity of terahertz channels and the architecture's flexibility, we present an enhanced algorithm that derives initial closed-form solutions from partial data, followed by comprehensive data fusion for refined estimation. This innovative approach substantially reduces computational complexity while preserving estimation accuracy.  
		
		\item \textbf{\textit{Theoretical Foundation:}} Addressing the fourth challenge, we develop a comprehensive CRLB  analysis for pilot-based DOA estimation under the designed HDS architecture. Through rigorous performance analysis and comparison across different architectures/connections via CRLB, we provide fundamental theoretical insights that guide the flexible design of various HDS architectures.
		
		\item \textbf{\textit{Simulation Validation:}} Extensive simulations and detailed computational complexity analyses demonstrate that our proposed approach achieves an exceptional balance among hardware efficiency, design flexibility, estimation accuracy, and computational complexity. These comprehensive results support the   applicability of our approach in THz UM-MIMO systems.
		
	\end{itemize}
	
	The remainder of this article is as follows: the system model for the THz UM-MIMO systems with the designed HDS architecture is formulated in Section II. The proposed 2-D DOA estimation algorithm   is presented in detail in Section III, while theoretical analysis and the corresponding CRLB are derived in Section IV. Numerical simulation results in various array configurations are provided in Section V, and conclusions are drawn in Section VI.

	\section{System Overview}
	In this section, we first describe our HDS architecture, and then introduce the THz UM-MIMO channel and system model.
	\begin{figure*}[!t]
		\centering
		\includegraphics[width=7.2in]{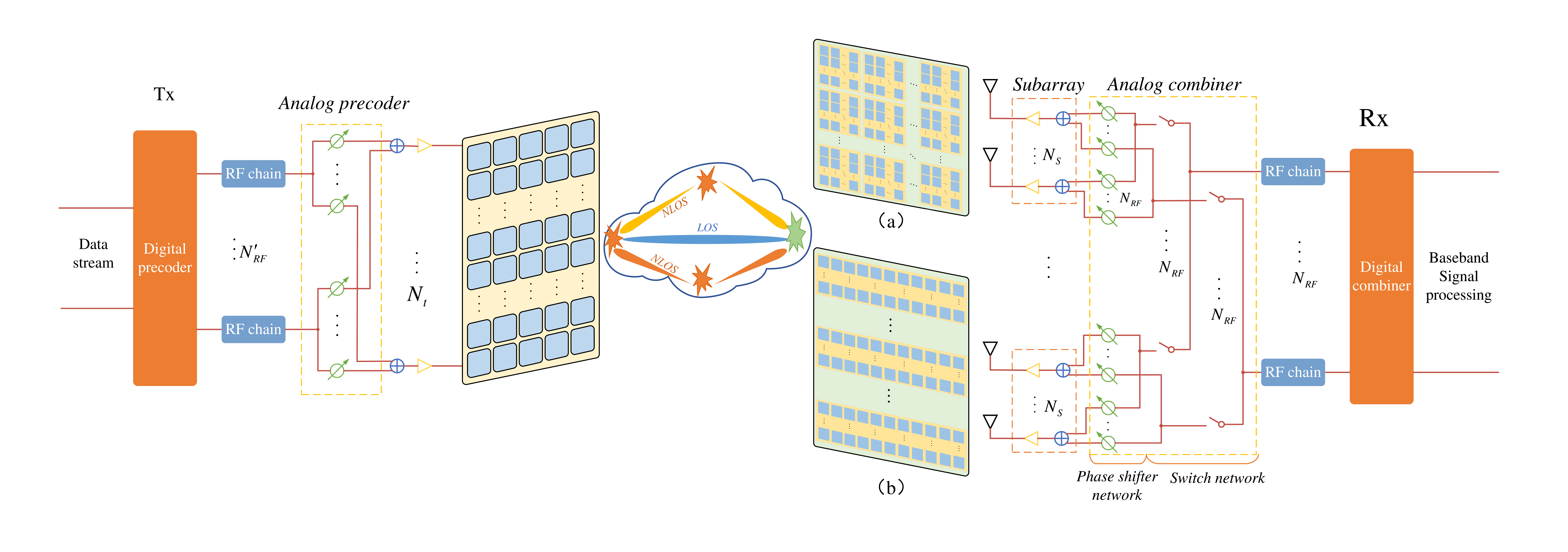}
		\caption{The adopted THz UM-MIMO system model, with HFC architecture at transmitter and HDS architecture (a. Existing  architecture; b. Proposed architecture.) at receiver.}
		\label{Fig1} 
	\end{figure*}
	\subsection{HDS Architecture}
	To fully harness the benefits of the HDS architecture and enable efficient DOA estimation, we propose a novel HDS architecture for the receiver (Rx), as illustrated in Fig.~\ref{Fig1}. In this architecture, the fundamental unit of the antenna array is expanded from a single antenna to a modular antenna group. The entire antenna array is partitioned into $N_\mathrm{RF}$ subarrays, corresponding to the number of RF chains, with each subarray comprising $N_s=N_r/ N_\mathrm{RF}$ antennas, where $N_r$ and $N_\mathrm{RF}$ represent the numbers of receivinge antennas and RF chains, respectively, and $N_\mathrm{RF}\ll N_r$. This modular division of the array is distinct from the connection/structure presented in \cite{r32}, as we have developed a unique design that explores the correlation between the received data from different RF chains.
	
	The subarray partitioning scheme is designed to simplify the 2-D DOA estimation problem into a 1-D one. For a 2-D uniform planar array (UPA), the subarrays are uniformly divided along one dimension while maintaining the same number of antennas as the original array in the other dimension. This ensures that the phase differences between subarrays are solely dependent on the 1-D DOAs, either azimuth or
	elevation angles.  By treating each subarray as an equivalent single antenna, the original 2-D planar array is transformed into an augmented uniform linear array, where the inter-subarray phase differences correspond to the antenna spacing. These phase differences are jointly determined by the number of antennas in the partitioning dimension and the number of RF chains. The proposed HDS architecture offers significant advantages for efficient DOA estimation by simplifying the 2-D problem to a 1-D one and leveraging the inter-subarray phase differences.

	A switch network exists between the RF chains and the subarrays, enabling intelligent adjustment of connections and realization of the functions mentioned above. Each RF chain can connect to any subarray through switches, with a total number of switches $N_\mathrm{RF}^{2}$. Let $0<\rho\leq1$  represent the proportion of closed switches in the system. When $\rho = 1/N_\mathrm{RF}$, the HDS architecture reduces to the HS architecture. Conversely, when $\rho = 1$, it evolves into the HFC architecture.  For intermediate values of $\rho$, the architecture provides a trade-off between hardware complexity and connection flexibility. For instance, when $\rho$ is closer to $1/N_\mathrm{RF}$,  the system minimizes RF chain utilization at the cost of reduced beamforming accuracy. On the other hand, as $\rho$ approaches 1, the system achieves higher performance with increased hardware overhead. This adaptability makes the HDS architecture suitable for diverse application scenarios, such as energy-efficient deployments or high-precision beamforming tasks. To maintain generality, the transmitter (Tx) is configured with a HFC architecture, and the beamforming technology is used to send signals through different paths to the Rx. The signals received by the antennas undergo a combination process through both analog combiner $\textbf{W}_\mathrm{A}\in\mathbb{C}^{N_r\times N_\mathrm{RF}}$ and digital combiner $\textbf{W}_\mathrm{D}\in\mathbb{C}^{N_\mathrm{RF}\times N_\mathrm{RF}}$.
	
	
	According to our HDS architecture in Fig.~\ref{Fig1}, the analog combining matrix $\mathbf{W}_\mathrm{A} $ at the Rx is represented as
	\begin{equation}\label{1}
		\mathbf{W}_\mathrm{A} = \mathbf{M}_{\mathrm{S}} \odot\mathbf{M}_{\mathrm{P}},
	\end{equation}
	where $\mathbf{M}_{\mathrm{S}} \in \{0,1\}^{{N_r\times N_\mathrm{RF}}}$ denotes the binary switch network
	matrix and $\mathbf{M}_{\mathrm{P}} \in \mathbb{C}^{{N_r\times N_\mathrm{RF}}}$ the phase shift network matrix.
	The specific structure of $\mathbf{M}_{\mathrm{S}}$ has the form of:
	\begin{equation}\label{2}
		\mathbf{M}_{\mathrm{S}} =  \begin{bmatrix}
			\mathbf{m}_{1,1} & \mathbf{m}_{1,2} & \cdots & \mathbf{m}_{1,N_\mathrm{RF}} \\
			\mathbf{m}_{2,1} & \mathbf{m}_{2,2} & \cdots & \mathbf{m}_{2,N_\mathrm{RF}} \\
			\vdots & \vdots & \ddots & \vdots \\
			\mathbf{m}_{N_\mathrm{RF},1} & \mathbf{m}_{N_\mathrm{RF},2} & \cdots & \mathbf{m}_{N_\mathrm{RF},N_\mathrm{RF}} \\
		\end{bmatrix},
	\end{equation}
	where $\mathbf{m}_{p,q}\in \{0,1\}^{N_s\times1}$ represents the connection status between the $p$-th subarray and the $q$-th RF chain. Specifically, when the switch is closed, $\mathbf{m}_{p,q} = \mathbf{1}_{N_s}$; otherwise, $\mathbf{m}_{p,q}=\mathbf{0}_{N_s}$.
	The phase shifter network matrix $\mathbf{M}_{\mathrm{P}}$ adheres to a constant modulus constraint, where the modulus is $1/\sqrt{N_r}$. In this work, the design of  $\mathbf{W}_A$ is crucial for achieving an efficient and flexible HDS architecture. The key considerations include determining the optimal connection pattern between the RF chains and subarrays in $\mathbf{M}_S$ to balance system performance and hardware complexity, and designing the phase shift values in $\mathbf{M}_P$ to ensure high DOA estimation accuracy while minimizing the impact of phase errors. The specific structure of $\mathbf{M}_S$ is chosen to facilitate the subsequent DOA estimation process and enable a flexible connection scheme between the RF chains and subarrays. 
	With the above definitions, we   rewrite $\mathbf{W}_\mathrm{A}$ as
	\begin{equation}\label{3}
		\mathbf{W}_\mathrm{A} =  \begin{bmatrix}
			\mathbf{w}_{1,1} & \mathbf{w}_{1,2} & \cdots & \mathbf{w}_{1,N_\mathrm{RF}} \\
			\mathbf{w}_{2,1} & \mathbf{w}_{2,2} & \cdots & \mathbf{w}_{2,N_\mathrm{RF}} \\
			\vdots & \vdots & \ddots & \vdots \\
			\mathbf{w}_{N_\mathrm{RF},1} & \mathbf{w}_{N_\mathrm{RF},2} & \cdots & \mathbf{w}_{N_\mathrm{RF},N_\mathrm{RF}} \\
		\end{bmatrix},
	\end{equation}
	where $\mathbf{w}_{p,q}$ represents the combined vector of the switch and phase shifters between the $p$-th subarray and the $q$-th RF chain. When the switch is closed and $\mathbf{m}_{p,q} = \mathbf{1}_{N_s}$,  $\mathbf{w}_{p,q}\in \mathbb{C}^{N_s\times1}$ is expressed as
	\begin{equation}\label{4}
		\mathbf{w}_{p,q} = \dfrac{1}{\sqrt{N_r}} \left[ e^{j\alpha_1}, e^{j\alpha_2}, \cdots, e^{j\alpha_{N_s}} \right],
	\end{equation}
	where $0 \leq \alpha_{n_s} \leq 2\pi $ denotes the phase shift coefficient, and when the switch is open, we have $\mathbf{w}_{p,q}=\mathbf{m}_{p,q}=\mathbf{0}_{N_s}$.
	
	The structure of $\mathbf{W}_\mathrm{A}$ provides an intuitive way to show the differences between various HAD architectures. In a HFC structure, there are no zero elements in $\mathbf{W}_\mathrm{A}$, whereas in a HS structure, $\mathbf{W}_\mathrm{A}$ is a block diagonal matrix. When the number of antennas far exceeds the number of RF chains, $\mathbf{W}_\mathrm{A}$ becomes sparse. The number of zero elements in $\mathbf{W}_\mathrm{A}$ captures the trade-off between system power consumption and received information. In the HDS structure, $\mathbf{W}_\mathrm{A}$ is flexibly controlled by adjusting switches, which means that even when considering practical hardware deployments,  a balance between system power consumption and performance can still be achieved.
	
	\subsection{Channel Model}
	The $N_r\times N_t $ THz UM-MIMO channel matrix is expressed as \cite{r32,r33,r34}
	\begin{equation}\label{5}
		\mathbf{H} = \sum_{\ell=1}^{L} \gamma_{\ell} \mathbf{a}_r(\theta_{\ell,r}, \phi_{\ell,r}) \mathbf{a}^H_t(\theta_{\ell,t}, \phi_{\ell,t}),
	\end{equation}
	where $L$ represents the number of multipaths in the THz UM-MIMO channel, which normally consists of one line-of-sight (LOS) path and several non-line-of-sight (NLOS) paths. Moreover, $\gamma_{\ell}$ represents the complex gain of the $\ell$-th path, which varies with the carrier frequency, propagation distance  and transmission path characteristics. In addition,  $\mathbf{a}_r(\theta_{\ell,r}, \phi_{\ell,r})\in \mathbb{C}^{N_r\times 1}$ and $\mathbf{a}_t(\theta_{\ell,t}, \phi_{\ell,t})\in \mathbb{C}^{N_t\times 1}$ represent the array steering vectors at the Rx and Tx, and the angle pairs $ \left( \theta_{\ell,r}, \phi_{\ell,r} \right)$ and $(\theta_{\ell,t}, \phi_{\ell,t})$ represent the DOA and direction-of-departure (DOD) of the $\ell$-th path, respectively.  $N_r$ and $N_t$ represent  the number of receiving and transmitting antennas. The parameters $\theta$ and $\phi$ represent the azimuth DOA and elevation DOA, respectively. Without loss of generality, an UPA equipped with $N_{x}\times N_{z}$ antennas is considered, with $N_{x}$ denoting the number of antennas in the horizontal direction, and $N_{z}$ in the vertical direction. The unified form of $\mathbf{a}_r(\theta_{\ell,r}, \phi_{\ell,r})$ and $\mathbf{a}_t(\theta_{\ell,t}, \phi_{\ell,t})$ is:
	\begin{multline}\label{6}
		{\mathbf{a}(\theta, \phi)=}[1,\cdots,e^{j2\pi d(n_x\sin\theta\cos\phi+n_z\sin\phi)/\lambda}, \\
		\cdots, e^{j2\pi d((N_{x}-1)\sin\theta\cos\phi+(N_{z}-1)\sin\phi)/\lambda}{]^{T}},
	\end{multline}
	where $\lambda$ denotes the carrier wavelength and $d = \lambda/2$ the inter-element spacing.
	
	Due to limited size of the UPA array and small ratio between the bandwidth and carrier frequency, the spatial selectivity effects are negligible in our considered scenario, which means that the channel matrix can be rearranged into a more compact matrix form as:
	\begin{equation}\label{7}
		\mathbf{H}=\mathbf{A}_r(\theta_r, \phi_r) \mathbf{\Lambda} \mathbf{A}_t^{H}(\theta_t, \phi_t).
	\end{equation}
	Here, $\mathbf{A}_r(\theta_r, \phi_r) \in\mathbb{C}^{ N_r\times L}$ and $\mathbf{A}_t(\theta_t, \phi_t)\in\mathbb{C}^{ N_t\times L}$ respectively denote the array manifold matrices at Rx and Tx, given by
	\begin{equation}\label{8}
		\mathbf{A}_r(\theta_r, \phi_r)=\left[\mathbf{a}_r(\theta_{1,r}, \phi_{1,r}),\cdots,\mathbf{a}_r(\theta_{L,r}, \phi_{L,r})\right],
	\end{equation}
	\begin{equation}\label{9}
		\mathbf{A}_t(\theta_t, \phi_t)=\left[\mathbf{a}_t(\theta_{1,t}, \phi_{1,t}),\cdots,\mathbf{a}_t(\theta_{L,t}, \phi_{L,t})\right],
	\end{equation}
	and $\mathbf{\Lambda}= \mathrm{diag}\{\gamma_1,\cdots,\gamma_L\}\in\mathbb{C}^{ L\times L}$ consists of the path gains of complex multipath channel \footnote{Our research is centered on low-complexity DOA estimation for THz UM-MIMO systems. For generality, we assume that the DOA remains constant within a single time slot but varies among different time slots. For DOA estimation, we depend solely on data from the first time slot, and thus, we do not dynamically model the channel in this study.}.

	\subsection{System Model}
	According to the above analysis, the system model of a THz UM-MIMO system configured with HDS connections is described as
	\begin{equation}\label{10}
		\mathbf{\Upsilon} = \mathbf{W}^{H}\mathbf{H}\mathbf{F}\mathbf{P}+\mathbf{W}^{{H}}\mathbf{N},
	\end{equation}
	where $\mathbf{P}\in\mathbb{C}^{N_a\times N_p}$ represents   the transmitted pilot symbol matrix, with $N_p$ ($\geq N_a$) and $N_a$ indicating the length of the pilot symbol and the number of data streams, respectively. For simplicity, we set $N_p=N_a$. The pilot symbols are orthogonal to each other, satisfying $\mathbf{P}\mathbf{P}^{H}=\mathbf{I}_{N_a}$. Here,  $\mathbf{F}=\mathbf{F}_\mathrm{A}\mathbf{F}_\mathrm{D}\in \mathbb{C}^{N_t\times N_a}$ represents the precoding matrix at the Tx, with $\mathbf{F}_\mathrm{D}$ and $\mathbf{F}_\mathrm{A}$ being the digital precoding matrix and the analog precoding matrix, respectively. We let $\mathbf{W}=\mathbf{W}_\mathrm{A}\mathbf{W}_\mathrm{D}\in \mathbb{C}^{N_r\times N_\mathrm{RF}}$ denote the overall combining matrix for the HDS architecture at the Rx. For ease of implementation and without loss of generality,  we design $\mathbf{F}_\mathrm{D}$ and $\mathbf{W}_\mathrm{D}$ as identity matrices, i.e., $\textbf{F}_\mathrm{D}=\textbf{I}_{N_a}$, $\textbf{W}_\mathrm{D}=\textbf{I}_{N_\mathrm{RF}}$. Furthermore, the additive noise $\mathbf{N}\in\mathbb{C}^{{N_r\times N_p}}$ is assumed to be white Gaussian distributed, each row obeys $\mathbf{n}\sim\mathcal{CN}(0, \sigma_n^{2} \mathbf{I}_{N_r})$.
	
	Multiplying $\mathbf{\Upsilon}$ by $\mathbf{P}^{H}$ at the Rx yields
	\begin{equation}\label{11}
		\mathbf{Y} = \mathbf{W}^{{H}}\mathbf{H}\mathbf{F}\mathbf{P}\mathbf{P}^{{H}}+\mathbf{W}^{{H}}\mathbf{N}\mathbf{P}^{{H}} \\
		= \mathbf{W}_{\mathrm{A}}^{{H}}\mathbf{H}\mathbf{F}_{\mathrm{A}}+\mathbf{\bar N},
	\end{equation}
	where $\mathbf{\bar N}=\mathbf{W}_{\mathrm{A}}^{H}\mathbf{N}\mathbf{P}^{H}\in\mathbb{C}^{{N_\mathrm{RF}\times N_a}}$ denotes the transformed additive noise.

	Substituting  \eqref{7} into \eqref{11}  yields:
	\begin{equation}\label{12}
		\mathbf{Y} = \mathbf{W}_{\mathrm{A}}^{H}\mathbf{A}_r(\theta_r, \phi_r)\mathbf{S}+\mathbf{\bar N},
	\end{equation}
	where $\mathbf{S}=\mathbf{\Lambda}\mathbf{A}_t^{{H}}(\theta_t, \phi_t)\mathbf{F}_{\mathrm{A}}\in\mathbb{C}^{L\times N_a}$ is defined as the equivalent signal matrix. Considering the structural characteristics of existing analog precoding matrix $\mathbf{F}_{\mathrm{A}}$, $\mathbf{\Lambda}$ and $\mathbf{A}_t(\theta_t, \phi_t)$ in (11), it is found that $\mathbf{S}$ possesses full row rank, which equals $L$. During the training procedure, the phase shift coefficient $\alpha_{n_{s}}$ in $\textbf{W}_\mathrm{A}$ is changed to generate different beams. Then, the Rx collects several channel observations to conduct DOA estimation. In this paper, we  estimate the unknown DOA parameters, including the azimuth angles $\theta_{l,r}$ and elevation angles $\phi_{l,r}$, $l=1,\cdots,L$.  Although the DODs, $\theta_{l,t}, \phi_{l,t}$, $l= 1,\cdots,L$ and channel gains are also unknown, they can all be incorporated into the equivalent signal matrix $\mathbf{S}$. Since $\mathbf{S}$ maintains its full row rank property, these unknown parameters do not affect   DOA estimation using subspace based apporoach (such as MUSIC), as most of them only rely on the orthogonality between the receive array steering vector $\mathbf{A}_r(\theta_r, \phi_r)$ and the noise subspace.
 \section{2-D DOA Estimation}
	In this section, we first describe the observation reconstruction procedure, and then exploit it to achieve low-complexity 2-D DOA estimation with the RD-MUSIC and improved RD-MUSIC algorithms.
 \subsection{Observation Reconstruction}
	The observation reconstruction in THz UM-MIMO systems is designed to transform the distributed spatial-temporal measurements into a unified observation matrix for efficient DOA estimation. The reconstruction process consists of three key phases: data collection, signal integration within RF chains, and final observation matrix construction.
	
	\subsubsection{Data Collection}
	In THz wireless communication systems, the time slot length is on the millisecond scale, while the symbol length is on the picosecond level. This characteristic enables us to collect multiple observations within a single time slot. Specifically, we use $T$ randomly generated combining matrices to capture the omni-directional channel components. The $\tau$-th combining matrix at the Rx is denoted as $\mathbf{W}_{\tau}$, where $\mathbf{w}_{p,q,\tau}$ represents the joint vector for the $p$-th subarray and $q$-th RF-chain. Without loss of generality, the state of the switches in HDS is considered to be random, with each switch having a 50\% probability of being closed. In particular, in designing the switch matrix, we utilize all subarrays and RF chains. Specifically, each row and each column of the matrix $\mathbf{M}_{\mathrm{S}}$ in \eqref{2} contains at least one all-one  vector $\mathbf{m}_{p,q}$.

	\subsubsection{Signal Integration within RF chains} 
	During the signal reconstruction phase, the characteristics of HDS architecture is leveraged, where the basic unit of the array is transformed from individual antennas to subarrays. Due to the distinct connections between different RF chains and subarrays, we represent the received data from each RF chain separately and integrate them at the final stage. According to the form of the combining matrix and the channel matrix, the observation vector for the $q$-th RF-chain is:
	\begin{equation}\label{13}
		\vspace{-2mm}	\mathbf{y}_{q,\tau} =
		\sum_{p=1}^{N_\mathrm{RF}} \mathbf{w}_{p,q,\tau}^H \mathbf{A}_s \mathbf{\Psi}_p \mathbf{S} + \mathbf{\bar n}_{q,\tau},
	\end{equation}
	where $\mathbf{y}_{q,\tau}\in\mathbb{C}^{1\times N_a}$ represents the received signal of the $p$-th RF chain when the $\tau$-th pilot signal is transmitted, $\mathbf{\bar n}_{q,\tau}\in\mathbb{C}^{1\times N_a}$ the received noise vector, and $\mathbf{A}_s\in\mathbb{C}^{N_{s}\times L}$ the array manifold matrix of the first subarray, consisting of the first $N_{s}$ rows of the $\mathbf{A}_r(\theta_r, \phi_r)$. The phase shift matrix $\mathbf{\Psi}_p \in \mathbb{C}^{L \times L}$ that denotes the phase difference between the first subarray and the $p$-th subarray is expressed as
	\begin{equation}\label{14}
		\mathbf{\Psi}_p = \mathrm{diag} \left\{ e^{j 2 \pi \frac{d}{\lambda} \beta_{1,i}}, \ldots, e^{j 2 \pi \frac{d}{\lambda} \beta_{L,i}} \right\},
	\end{equation}
	where $\beta_{l,p} = (p-1) (N_z /N_\mathrm{RF}) \sin \phi_{l,r}$ \footnote{This decomposition of array manifold matrix reduces computational complexity while providing clearer spatial structure through explicit phase differences.}.
	
	Consequently, the matrix form of $\mathbf{y}_{q,\tau}$ with $T$ observations is given by
	\begin{equation}\label{15}
		\mathbf{\bar{Y}}_q = \begin{bmatrix} \mathbf{y}_{q,1}^T, \ldots, \mathbf{y}_{q,T}^T \end{bmatrix}^T = \mathbf{\bar{W}}^H_q\mathbf{A}_r \mathbf{S} + \tilde{\mathbf{N}}_q,
	\end{equation}
	where $\mathbf{\bar{Y}}_q\in\mathbb{C}^{T\times N_a}$, $\mathbf{\bar{W}}_q = \left[ \mathbf{\bar{w}}_{q,1}, \ldots, \mathbf{\bar{w}}_{q,T} \right] \in \mathbb{C}^{N_r \times T}$, $\mathbf{A}_r=\mathbf{A}_r(\theta_r, \phi_r)$, and
	$\tilde{\mathbf{N}}_q = \left[ \mathbf{\bar n}_{q,1}^T, \dots, \mathbf{\bar n}_{q,T}^T \right]^T \in \mathbb{C}^{T \times N_a}$ with $\mathbf{\bar{w}}_{q,\tau} \in \mathbb{C}^{N_r \times 1} $ representing the joint vector of the $q$-th RF-chain and the entire antenna array for the $\tau$-th combining matrix.
	
	\subsubsection{Final Observation Matrix Construction}
	By integrating the observations across all RF chains, we construct the final observation matrix as
	\begin{equation}\label{16}
		\mathbf{\tilde{Y}} = \begin{bmatrix} \mathbf{\bar{Y}}_1^{T}, \ldots, \mathbf{\bar{Y}}_{N_\mathrm{RF}}^{T} \end{bmatrix}^{T} = \mathbf{\tilde{W}}^{H}\mathbf{A}_r \mathbf{S} + \tilde{\mathbf{N}},
	\end{equation}
	where $\mathbf{\tilde{W}}=\begin{bmatrix}\mathbf{\bar{W}}_1,\dots,\mathbf{\bar{W}}_{N_\mathrm{RF}}\end{bmatrix}\in\mathbb{C}^{N_r\times N_\mathrm{RF}T}$ and $\tilde{\mathbf{N}}=\begin{bmatrix}\tilde{\mathbf{N}}_1^{T},\dots,\tilde{\mathbf{N}}_{N_\mathrm{RF}}^{T}\end{bmatrix}^{T}\in\mathbb{C}^{N_\mathrm{RF}T\times N_a}$. In what follows. it will be demonstrated how to exploit $\mathbf{\tilde{Y}}$ to achieve low-complexity and high-accuracy 2-D DOA estimation. 
	 \subsection{RD-MUSIC for 2-D DOA Estimation}
	To achieve high-accuracy 2-D DOA estimation without performing a 2-D spectral search, the RD-MUSIC based algorithm is applied. Given $\mathbf{\tilde{Y}}$, the sample covariance matrix (SCM) is calculated by
	\begin{equation}\label{17}
		\mathbf{R}_\mathrm{y} = \frac{1}{N_a} \mathbf{\tilde{Y}} \mathbf{\tilde Y}^H = \mathbf{\tilde{E}} \mathbf{R}_\mathrm{s}\mathbf{\tilde{E}}^H + \mathbf{R}_\mathrm{n},  \vspace{-2mm}
	\end{equation}
	where $\mathbf{R}_\mathrm{s}=\frac{1}{N_a} \mathbf{S} \mathbf{S}^H$ and $\mathbf{R}_\mathrm{n} = \frac{1}{N_a} \mathbf{\tilde{N}} \mathbf{\tilde{N}}^H$ represent the covariance matrix of the equivalent signal $\mathbf{S}$ and noise $\tilde{\mathbf{N}}$, respectively, while  $\mathbf{\tilde{E}}=\mathbf{\tilde{W}}^H\mathbf{A}_r$ denotes the augmented array manifold matrix.
	
	Performing eigenvalue decomposition (EVD) on $\mathbf{R}_\mathrm{y}$ yields
	\begin{equation}\label{18}
		\mathbf{R}_\mathrm{y} =\mathbf{U}_\mathrm{s} \mathbf{\Sigma}_\mathrm{s} \mathbf{U}_\mathrm{s}^{H} + \mathbf{U}_\mathrm{n} \mathbf{\Sigma}_\mathrm{n} \mathbf{U}_\mathrm{n}^{H},
	\end{equation}
	where $\mathbf{\Sigma}_\mathrm{s}\in \mathbb{C}^ {L\times L}$ and $\mathbf{\Sigma}_\mathrm{n}\in \mathbb{C}^{(N_\mathrm{RF}T-L)\times (N_\mathrm{RF}T-L)}$ are two diagonal matrices that consist of $L$ largest eigenvalues and remaining eigenvalues, respectively. $\mathbf{U}_\mathrm{s}\in \mathbb{C}^{N_\mathrm{RF}T\times L}$ and $\mathbf{U}_{\mathrm{n}} \in \mathbb{C}^{N_{\mathrm{RF}}T \times (N_{\mathrm{RF}}T - L)}$ are eigenvectors corresponding to $\mathbf{\Sigma}_\mathrm{s}$ and $\mathbf{\Sigma}_\mathrm{n}$, respectively.
	
	According to subspace theory, $\mathbf{U}_\mathrm{n}$ is orthogonal to the array manifold matrix, which implies that
	\begin{equation}\label{19}
		{{\bf{a}}^H}(\theta ,\phi ){{\bf{\tilde W}}}{{\bf{U}}_{\rm{n}}}{\bf{U}}_{\rm{n}}^H{{\bf{\tilde W}}^H\bf{a}}({\bf{\theta }},\phi ) = 0,
	\end{equation}
	for the true DOA pairs.
	
	For the reconstructed observation matrix $\mathbf{\tilde{Y}} \in\mathbb{C}^{N_\mathrm{RF}T\times N_a} $, it should be noted that $N_\mathrm{RF}T$ and $N_a$ are regarded as the virtual numbers of antennas and samples, respectively. In the general  THz systems, $N_\mathrm{RF}T$ is normally a large and comparable with $N_a$, i.e., large-scale and/or high-dimensional scenarios. Under such circumstances, the eigenvalues and eigenvectors obtained directly by EVD are not their true ones anymore, according to the GAT in \cite{r35,r36}, implying that the relationship in  \eqref{19} no longer holds rigorously.
	
	To address this issue and subsequently achieve robust DOA estimation, the phase compensation technique developed under the GAT framework is applied. 
	To achieve the bias-corrected estimators, we propose the following theorem.
	
	\emph{Theorem 1 (\cite{r37,r38}):} Suppose that $\hat \lambda _\ell$ and ${\bf{\hat u}_\ell}$ are eigenvalues and eigenvectors corresponding to the signal subspace of $\mathbf{R}_\mathrm{y}$, respectively, provided that $N_\mathrm{RF}T$ and $N_a$ are large, namely,    $N_\mathrm{RF}T\to\infty$, $N_a\to\infty$ and $c=N_\mathrm{RF}T/N_a<\infty$; $\varepsilon_\ell$ and ${\bf v}_\ell$ are true and/or unbiased eigenvalues of $\mathbf{\tilde{E}} \mathbf{R}_\mathrm{s}\mathbf{\tilde{E}}^H$, respectively, $\ell=1,\ldots,L$. The following convergence results hold almost surely
	\begin{equation}\label{20}
		\left[ {{{\hat \lambda }_\ell }} \right]_{\ell  = 1}^L \xrightarrow{a.s.} \left\{ {\begin{array}{*{20}{c}}
				{\frac{{(\bar \sigma _n^2 + {\varepsilon _\ell })(\bar \sigma _n^2c + {\varepsilon _\ell })}}{{{\varepsilon _\ell }}},}&{{\varepsilon _\ell } > \bar \sigma _n^2\sqrt c },\\
				{\bar \sigma _n^2{{(1 + \sqrt c )}^2},}&\text {otherwise},
		\end{array}} \right.
	\end{equation}
	\begin{equation}\label{21}
		\frac{N_\mathrm{RF}T-L}{\bar\sigma_{n}^{2} \sqrt{2 c}}\left(\hat{\sigma}_{n}^{2}-\bar \sigma_{n}^{2}\right)+b\left(\bar\sigma_{n}^{2}\right) \xrightarrow{D} \mathcal{N}(0,1),
	\end{equation}
	\begin{equation}\label{22}
		\left\langle\hat{\mathbf{u}}_\ell, \mathbf{v}_\ell\right\rangle^{2} \xrightarrow{\text { a.s. }}\left\{\begin{array}{ll}\frac{\varepsilon_\ell^{2}-\bar \sigma_{n}^{4}c}{\varepsilon_\ell\left(\varepsilon_\ell+\bar \sigma_{n}^{2} c\right)}, & \varepsilon_\ell \geq \bar \sigma_{n}^{2} \sqrt{c}, \\ 0, & \text {otherwise},\end{array}\right.
	\end{equation}
	where ${\bar \sigma _n^2}$ represents the true variance of noise $\tilde{\mathbf{N}}$ in (16), $\hat{\sigma}_{n}^{2}$ the maximum likelihood estimate of ${\bar \sigma _n^2}$ and
	\begin{equation}
	b(\bar \sigma _n^2) = \sqrt {c/2} (L + \bar \sigma _n^2\sum\nolimits_{\ell  = 1}^L {1/{\varepsilon _\ell }} ).
	\end{equation}
	
		\begin{proof}
	Refer to \cite{r37,r38}.
	\end{proof} 
	
	Theorem 1 indicates that both eigenvalues and eigenvectors obtained directly through EVD exhibit deviations.  Based on   \eqref{20}-\eqref{22} from Theorem 1, it is further determined that the following bias-corrected estimators with respect to ${\bar \sigma _n^2}$, $\varepsilon_\ell$ and ${\bf v}_\ell$ are, respectively:
	\begin{equation}\label{23}
		\tilde{\sigma}_{n}^{2}=\hat{\sigma}_{n}^{2}+\frac{b\left(\hat{\sigma}_{n}^{2}\right)}{N_\mathrm{RF}T-L} \hat{\sigma}_{n}^{2} \sqrt{2 c},
	\end{equation}
	\begin{equation}\label{24}
		\tilde{\varepsilon}_\ell=\frac{1}{2}\{\hat{\lambda}_\ell-\hat{\sigma}_{n}^{2}(1+c)+\sqrt{[\hat{\lambda}_\ell-\hat{\sigma}_{n}^{2}(1+c)]^{2}-4 \hat{\sigma}_{n}^{2} c}\},
	\end{equation}
	\begin{equation}\label{25}
		{{{\bf{\tilde v}}}_\ell} = {\left\langle {{{{\bf{\hat u}}}_\ell},{{\bf{v}}_\ell}} \right\rangle ^{ - 1}}{{{\bf{\hat u}}}_\ell} = \vartheta _\ell^{ - 1}{{{\bf{\hat u}}}_\ell},
	\end{equation}
	where ${\vartheta _\ell} = \left\langle {{{{\bf{\hat u}}}_k},{{\bf{v}}_\ell}} \right\rangle $ denotes the phase transformation factor between ${{{{\bf{\hat u}}}_\ell}}$ and ${{{\bf{v}}_\ell}}$.
	
	By employing   \eqref{23}-\eqref{25} and the fact ${{\bf{U}}_{\rm{n}}}{\bf{U}}_{\rm{n}}^H = {{\bf{I}}_{{N_{{\rm{RF}}}}T}} - {{\bf{U}}_{\rm{s}}}{\bf{U}}_{\rm{s}}^H$, we can correct the estimate of ${{\bf{U}}_{\rm{n}}}{\bf{U}}_{\rm{n}}^H$ as
	\begin{equation}\label{26}
		{{{\bf{\bar U}}}_{\rm{n}}}{\bf{\bar U}}_{\rm{n}}^H = {{\bf{I}}_{{N_{{\rm{RF}}}}T}} - \sum\nolimits_{\ell  = 1}^L {{{\left({\tilde \vartheta _\ell ^2} \right)}^{ - 1}}{\bf{\hat u}_\ell }{\bf{\hat u}_\ell} ^H},
	\end{equation}
	where $\tilde{\vartheta}_\ell^{2}=\frac{\tilde{\varepsilon}_\ell^{2}-\tilde{\sigma}_{n}^{4} c}{\tilde{\varepsilon}_\ell\left(\tilde{\varepsilon}_\ell+\tilde{\sigma}_{n}^{2} c\right)}$.
 
	To simplify the estimator, we apply dimensionality reduction to transform the 2-D search problem into two separate 1-D search problems.
	
	Specifically,  define $u=\sin \phi$ and $v=\sin \theta \cos \phi$. Then the array steering vector in \eqref{6} is rewritten as $\mathbf{a}(\theta, \phi)=\mathbf{a}_z(u) \otimes \mathbf{a}_x(v)$, where $\mathbf{a}_z(u)= \begin{bmatrix} 1,e^{j2\pi du/\lambda},\ldots, e^{j2\pi(N_{z}-1) du/\lambda}\end{bmatrix}^T$ and $\mathbf{a}_x(v)= \begin{bmatrix} 1,e^{j2\pi dv/\lambda},\ldots, e^{j2\pi(N_{x}-1) dv/\lambda}\end{bmatrix}^T$. Subsequently, we construct
	\begin{align*}
		\mathbf{V}(u, v) &= \left[ \mathbf{a}_z(u) \otimes \mathbf{a}_x(v) \right]^H \mathbf{\tilde{W}} \mathbf{\bar U}_\mathrm{n} \mathbf{\bar U}^H_\mathrm{n} \mathbf{\tilde{W}}^H \left[ \mathbf{a}_z(u) \otimes \mathbf{a}_x(v) \right] \\
		&=\mathbf{a}^H_x(v) \left[ \mathbf{a}_z(u) \otimes \mathbf{I}_{N_x} \right]^H \mathbf{E}_\mathrm{n} \mathbf{E}^H_\mathrm{n} \left[ \mathbf{a}_z(u) \otimes \mathbf{I}_{N_x} \right]\mathbf{a}_x(v)\notag \\
		&=\mathbf{a}^H_x(v)\mathbf{Q}(u)\mathbf{a}_x(v),
	\end{align*}
	where $\mathbf{E}_\mathrm{n}=\mathbf{\tilde{W}} \mathbf{\bar U}_\mathrm{n}$ represents the augmented noise subspace, and $\mathbf{Q}(u)=\left[ \mathbf{a}_z(u) \otimes \mathbf{I}_{N_x} \right]^H \mathbf{E}_\mathrm{n} \mathbf{E}^H_\mathrm{n} \left[ \mathbf{a}_z(u) \otimes \mathbf{I}_{N_x} \right]$.
 
	With $\mathbf{V}(u, v)$ and the auxiliary constraint of $\mathbf{d}^H_1\mathbf{a}_x(v)=1$, $u$ and $v$ are estimated by solving the following quadratic optimization problem
	\begin{equation}\label{27}
		\left\{ {\hat u,\hat v} \right\} = \arg \mathop {\min }\limits_{u,v} {\bf{a}}_x^H(v){\bf{Q}}(u){{\bf{a}}_x}(v){\;}s{\rm{. }}t{\rm{. }}\;{\bf{d}}_1^H{{\bf{a}}_x}(v) = 1
	\end{equation}
	where $\mathbf{d}_1=\begin{bmatrix}1,0,\ldots,0 \end{bmatrix}^{T}\in \mathbb{R}^{N_x \times 1}$
	
	By applying the Lagrange multiplier method, we construct the Lagrangian   associated with \eqref{20}  as
	\begin{equation}\label{28}
		L(\theta, \phi) = \mathbf{a}^H_x(v) \mathbf{Q}(u) \mathbf{a}_x(v) - \varsigma \left( \mathbf{d}_1^{H} \mathbf{a}_x(v) - 1 \right)
	\end{equation}
	where $\varsigma>0$  is the Lagrange multiplier. Setting the partial derivative of  \eqref{28} with respect to $\mathbf{a}_x(v)$ to zero   yields
	\begin{equation}\label{29}
		\mathbf{a}_x(v)=\frac{\mathbf{Q}^{-1}(u)\mathbf{d}_1}{\mathbf{d}^H_1 \mathbf{Q}^{-1}(u) \mathbf{d}_1}.
	\end{equation}
Substituting \eqref{29}  back to the constraint   \eqref{27}, $u$ is estimated from:
	\begin{equation}\label{30}
		\hat u = \arg \mathop {\min }\limits_u \frac{1}{{{\bf{d}}_1^H{{\bf{Q}}^{ - 1}}(u){{\bf{d}}_1}}}.
	\end{equation}
	
	Performing a 1-D spectral search to find the $L$ values of $\left\{ {{{\hat u}_\ell}} \right\}_{\ell = 1}^L$ that minimize the cost function  \eqref{30}, the elevation DOAs are estimated as
	\begin{equation}\label{31}
		{{\hat \phi }_\ell} = \arcsin \left( {{{\hat u}_\ell}} \right) \cdot  180/\pi ,\;\ell = 1, \ldots ,L.
	\end{equation}
 With available $\{{{\hat \phi }_\ell}\}_{\ell = 1}^L$, $\mathbf{V}(u, v)$ reduces to $\mathbf{V}(v)$. Then, by conducting 1-D spectral search on $\mathbf{V}(v)$  again with variable $v$ to obtain $L$ values of $\left\{ {{{\hat v}_\ell}} \right\}_{\ell = 1}^L$ that minimize $\mathbf{V}(v)$, the estimates of azimuth DOAs are obtained via
	\begin{equation}\label{32}
		{{\hat \theta }_\ell } = \arcsin \left( {\frac{{{{\hat v}_\ell }}}{{\cos ({{\hat \phi }_\ell })}}} \right) * 180/\pi ,\ell  = 1, \ldots ,L.
	\end{equation}
	
	The RD-MUSIC procedure for 2-D DOA estimation  is summarized in \textbf{Algorithm 1}. 
	\begin{algorithm}[!t]
		\caption{RD-MUSIC for the HDS Architecture}
		\LinesNumbered
		\KwIn{Array output matrix $\mathbf{\Upsilon}$, pilot symbol matrix $\mathbf{P}$ and combining matrix ${\bf{\tilde W}}$.}
		\KwOut{$(\hat \theta _\ell,\hat \phi _\ell) $}
		Construct an expanded array output matrix $\mathbf{\tilde{Y}}$ within $T$ observations.\\
		Calculate the SCM through $\mathbf{R}_\mathrm{y} = \frac{1}{N_a} \mathbf{\tilde{Y}} \mathbf{\tilde Y}^H$.\\
		Perform EVD on $\mathbf{R}_\mathrm{y}$ to obtain noise subspace matrix $\mathbf{U}_\mathrm{n}$.\\
		Construct bias-corrected estimators with respect to ${\bar \sigma _n^2}$ and $\varepsilon_\ell$  based on (23), (24).\\
		Calculate ${{{\bf{\bar U}}}_{\rm{n}}}{\bf{\bar U}}_{\rm{n}}^H$ according to (26).\\
		
		\For{ $\phi\in$ search grid}{
			Construct $\mathbf{Q}(u)$ utilizing ${\bf{\tilde W}}$ and ${{{\bf{\bar U}}}_{\rm{n}}}{\bf{\bar U}}_{\rm{n}}^H$, and then compute its inverse.\\
			Return $\hat {u}_\ell$, $\ell=1,\ldots, L$, according to (23);\\
			Compute elevation DOA estimates using
			${{\hat \phi }_\ell} = \arcsin \left( {{{\hat u}_\ell}} \right) * 180/\pi ,\;\ell = 1, \ldots ,L$.}
		\For{ $\theta\in$ search grid}{
			
			Return $\hat {v}_\ell$, $\ell=1,\ldots, L$, according to $\hat v = \arg \mathop {\min }\limits_v {\bf{a}}_x^H(v){\bf{Q}}(u){{\bf{a}}_x}(v)$.\\
			Compute azimuth DOA estimates using ${{\hat \theta }_\ell } = \arcsin \left( {\frac{{{{\hat v}_\ell }}}{{\cos ({{\hat \phi }_\ell })}}} \right) * 180/\pi ,\ell  = 1, \ldots ,L.$ \\}
		Return 2-D DOA pairs $\{{(\hat \theta _\ell,\hat \phi _\ell)}\}_{\ell = 1}^L$. 
	\end{algorithm} \vspace{-2mm}
	\vspace{-2mm}	\subsection{Improved RD-MUSIC for 2-D DOA Estimation}
	Although RD-MUSIC avoids the need for a full 2-D spectral search, it still requires two 1-D spectral searches across the entire angular domain. In practice, due to the high transmission loss of THz waves, energy is typically concentrated in only a few transmission channels, resulting in significant sparsity in the angular domain. This sparsity allows for a substantial reduction in the search space.  Below, we explain how to leverage this sparsity to further reduce complexity of the estimator.
	
	\subsubsection{Special HDS Architecture}
	Furthermore, leveraging our   HDS architecture, we   select two RF chains for specialized configuration, enabling initial DOA estimation through \textbf{\textit{closed-form}} solutions and further narrowing the search space. Specifically, the first RF chain is connected to the first $N_\mathrm{RF}-1$ subarrays, while the second RF chain is connected to the last $N_\mathrm{RF}-1$ subarrays. This configuration is easily realized by adjusting the switch network.
	In addition, we ensure that the phase shift coefficients generated by the phase shifters connected to the second RF chain are identical to those of the first RF chain, which means that
	\begin{equation}\label{33}
		\mathbf{w}_{i,1}=\mathbf{w}_{i+1,2},\;\; i=1,\ldots,N_\mathrm{RF}-1.
	\end{equation}
	
	\subsubsection{Elevation Angle Information}
	Under this design, $\mathbf{W}_\mathrm{A}$ remains a full-rank matrix with a rank of $N_\mathrm{RF}$, but a constant phase difference exists between the received signals of the second and the first RF chains. Let $\mathbf{w}_1$ and $\mathbf{w}_2$ denote the first and second columns of $\mathbf{W}_\mathrm{A}$, respectively. Based on the connection configuration described in  \eqref{33}, we have
	\begin{equation}\label{34}
		\mathbf{w}_2^{H}\mathbf{A}_r =\mathbf{w}_1^{H}\mathbf{A}_r\boldsymbol{\Psi},
	\end{equation}
	\begin{equation}\label{35}
		\boldsymbol{\Psi}={\rm{diag}}\left\{ {e^{j2\pi \frac{{d{N_z}}}{{\lambda {N_{{\rm{RF}}}}}}\sin {\phi _1}}}, \ldots ,{e^{j2\pi \frac{{d{N_z}}}{{\lambda {N_{{\rm{RF}}}}}}\sin {\phi _L}}}\right\},
	\end{equation}
	from which it is evident that the \textit{\textbf{elevation angle information}} of the received signal is extracted from $\boldsymbol{\Psi}$.
	\subsubsection{Closed-Form Solution}
	We further assume that the reconstructed observation matrix within $T$ pilots on the first and the second RF chains are $\mathbf{\bar Y}_1$ and $\mathbf{\bar Y}_2$, and $\mathbf{\tilde{N}}_1$ and $\mathbf{\tilde{N}}_2$ are their corresponding complex additive white Gaussian noise matrices, respectively. Then, we have
	\begin{equation}\label{36}
		\mathbf{\bar{Y}}_1 = \mathbf{\bar{W}}^H_1\mathbf{A}_r \mathbf{S} + \tilde{\mathbf{N}}_1  = \mathbf{\bar A}_1\mathbf{S} + \tilde{\mathbf{N}}_1,
	\end{equation}
	\begin{equation}\label{37}
		\mathbf{\bar{Y}}_2  = \mathbf{\bar{W}}^H_1\mathbf{A}_r \boldsymbol{\Psi} \mathbf{S} + \tilde{\mathbf{N}}_2 = \mathbf{\bar A}_1 \boldsymbol{\Psi}\mathbf{S} + \tilde{\mathbf{N}}_2,
	\end{equation}
	where $\mathbf{\bar A}_1=\mathbf{\bar{W}}^H_1\mathbf{A}_r\in \mathbb{C}^{T\times L}$ represents the augmented array covariance matrix with respect to the first RF chain. To make $\mathbf{\bar A}_1$ to be a full column rank matrix, $T>L$ is required.
	
The signal subspace spans the same space with array manifold matrix. That is, there exists a nonsingular matrix $\mathbf{T} \in \mathbb{C}^{L\times L}$, satisfying
	\begin{equation}\label{38}
		\mathbf{U}_\mathrm{s_1} = \mathbf{\bar A}_1 \mathbf{T}, \mathbf{U}_\mathrm{s_2}= \mathbf{\bar A}_1 \boldsymbol{\Psi} \mathbf{T},
	\end{equation}
	where $\mathbf{ U}_\mathrm{s_1} \in \mathbb{C}^{T\times L}$ and $\mathbf{ U}_\mathrm{s_2}\in \mathbb{C}^{T\times L}$ stand for the augmented signal subspace matrices associated with sample covariance matrices of $\mathbf{\bar{Y}}_1$ and $\mathbf{\bar{Y}}_2$, respectively.  It is necessary to point out that the augmented signal subspaces $\mathbf{U}_\mathrm{s_1}$ and $\mathbf{U}_\mathrm{s_2}$ are directly obtained from $\mathbf{U}_\mathrm{s}$ without performing additional EVD, which consist of the elements of the first $T$ rows and the rows from $T+1$ to $2T$, respectively. 
	
	Consequently, by performing EVD on ${\bf{U}}_\mathrm{s_1}^\dag {{\bf{U}}_\mathrm{s_2}}$ to obtain $L$ eigenvalues $\left\{ {{\beta _1}, \ldots ,{\beta _L}} \right\}$, the initial elevation DOAs are estimated as
	\begin{equation}\label{39}
		{\tilde{\phi}_\ell } = \arcsin \left( {\frac{{\angle \left[ {{\beta _\ell }} \right] \cdot \lambda {N_{{\rm{RF}}}}}}{{2\pi d{N_z}}}} \right),\ell  = 1, \ldots ,L.
	\end{equation}
An enhanced elevation DOA estimate ${\hat \phi _\ell }$ is obtained by conducting the RD-MUSIC algorithm around its initial estimate with a very small search space, which reduces  computational complexity while maintaining high accuracy.
	\begin{algorithm}[!t]
		\caption{IMRD-MUSIC for the HDS Architecture}
		\KwIn{Array output matrix $\mathbf{\Upsilon}$, pilot symbol matrix $\mathbf{P}$ and combining matrix ${\bf{\tilde W}}$.}
		\KwOut{$(\hat \theta _\ell,\hat \phi _\ell) $}
		Execute the first three steps of Algorithm 1 to obtain $\mathbf{R}_\mathrm{y}$.\\
		Perform EVD on $\mathbf{R}_\mathrm{y}$ to obtain noise subspace matrix $\mathbf{U}_\mathrm{n}$ and signal subspace matrix $\mathbf{U}_\mathrm{s}$. \\
		Extract $\mathbf{U}_\mathrm{s_1}$ and $\mathbf{U}_\mathrm{s_2}$ from $\mathbf{U}_\mathrm{s}$, then calculate ${\bf{U}}_\mathrm{s_1}^\dag {{\bf{U}}_\mathrm{s_2}}$.\\
		Perform EVD on ${\bf{U}}_\mathrm{s_1}^\dag {{\bf{U}}_\mathrm{s_2}}$ to yield $\boldsymbol{\Psi}$.\\
		Obtain initial elevation DOA estimates as ${\hat \phi _\ell } = \arcsin \left( {\frac{{\angle \left[ {{\beta _\ell }} \right] \cdot \lambda {N_{{\rm{RF}}}}}}{{2\pi d{N_z}}}} \right),\ell  = 1, \ldots ,L.$\\
		
		\For{each ${\phi}_\ell \in~small~search~grid~around~{\tilde{\phi}_\ell}$ }{
			Extract   steps 7-9 of Algorithm 1 to obtain enhanced elevation DOA estimate, denoted as ${{\hat \phi }_\ell}$.\\
		}
		\For{each $\hat{\phi}_\ell \in \{{{\hat \phi }_\ell}\}_{\ell = 1}^L$ }{
			Estimate $\mathbf{a}_x(v)$ as $\mathbf{\hat a}_x(v) = \frac{\mathbf{Q}^{-1}(\hat u)\mathbf{d}_1}{\mathbf{d}^H_1 \mathbf{Q}^{-1}(\hat u) \mathbf{d}_1}$.\\
			Extract the phase of $\mathbf{\hat a}_x(v)$ as $\mathbf{\hat{g}}_\ell = \angle[\mathbf{\hat a}_x(v)]$.\\
			Obtain   LS solution as ${{{\bf{\hat c}}}_\ell } = {{\bf{P}}^\dag }{{{\bf{\hat g}}}_\ell }$.\\
			Output the azimuth DOA estimate as ${{\hat \theta }_\ell } = \arcsin \left( {\frac{{{{\hat v}_\ell }}}{{\cos ({{\hat \phi }_\ell })}}} \right) \cdot 180/\pi ,\ell  = 1, \ldots ,L.$.\\
		}
		Return 2-D DOA pairs $\{{(\hat \theta _\ell,\hat \phi _\ell)}\}_{\ell = 1}^L$.
	\end{algorithm}
	
Employing ${\hat \phi _1},\ldots,{\hat \phi _L}$, the estimate $\mathbf{\hat a}_x(v)$ of $\mathbf{a}_x(v)$ is obtained based on \eqref{29}. It is noted that the phase of $\mathbf{a}_x(v)$ is expressed as $\mathbf{g}=\begin{bmatrix} 0,2\pi dv/\lambda,\ldots, 2\pi(N_{x}-1) dv/\lambda\end{bmatrix}^T$. Then, by extracting the phase of $\mathbf{\hat a}_x(v)$,  $v$ is finally determined by solving the following least squares (LS) optimization problem
	\begin{equation}\label{40}
		\min_{\mathbf{c}_\ell} \left\| \mathbf{P}\mathbf{c}_\ell - \mathbf{\hat{g}}_\ell \right\|_F^2,  \vspace{-2mm}
	\end{equation}
	where
	\begin{equation}\label{41}
		\mathbf{P} =
		\begin{bmatrix}
			1 & 1\dots  & 1 \\
			
			0 & 1 \dots &N_x-1
		\end{bmatrix}^T,  \vspace{-3mm}
	\end{equation}
	$\mathbf{c}_\ell=\begin{bmatrix} c_{\ell_0}, v_\ell\end{bmatrix}^T\in\mathbb{R}^{2\times1}$ is an unknown parameter vector  with  $c_{\ell_0}$ being the  parameter error.
	
	The LS solution of  \eqref{40} is ${{{\bf{\hat c}}}_\ell } = {{\bf{P}}^\dag }{{{\bf{\hat g}}}_\ell }$, $\ell=1,\ldots, L$. By utilizing the second element in ${{{\bf{\hat c}}}_\ell }$ and   \eqref{32}, the azimuth DOA of the $\ell$-th path is estimated as ${{\hat \theta }_\ell }$, and ${{\hat \theta }_\ell }$ and ${\hat \phi _\ell}$ are automatically paired. The improved RD-MUSIC method, referred to as IMRD-MUSIC,  for 2-D DOA estimation is summarized in \textbf{Algorithm 2}.
	\begin{table*}[t]
		\centering
		\captionsetup{justification=centering, labelsep=newline}
		\caption{COMPUTATIONAL COMPLEXITY COMPARISON}
		\renewcommand{\arraystretch}{}
		\begin{tabular}{c c c c c}
			\toprule
			\textbf{Algorithm} & \textbf{SCM Calculation} & \textbf{EVD} & \textbf{Offline Calculation of ${{\bf{U}}_{\rm{n}}}{\bf{U}}_{\rm{n}}^H$/${{\bf{E}}_{\rm{n}}}{\bf{E}}_{\rm{n}}^H$} & \textbf{Spectral Search} \\ \midrule
			\vspace{0pt}
			FD-2D-MUSIC & $\mathcal{O}(N_r^2TN_a)$ & $\mathcal{O}(N_r^3)$ & $\mathcal{O}\Big(N_r^2(N_r-L)\Big)$ & $\mathcal{O}(N_r^2{n_\theta}{n_\phi})$ \\
			HDS-2D-MUSIC & $\mathcal{O}({N_{RF}^2}{T^2}N_a)$ & $\mathcal{O}({N_{RF}^3}T^3)$ & $\mathcal{O}\Big({N_r^2}(N_{RF}T-L)\Big)$ & $\mathcal{O}({N^2_{r}}{n_\theta}{n_\phi})$ \\
			HDS-RD-MUSIC & $\mathcal{O}({N_{RF}^2}{T^2}N_a)$ & $\mathcal{O}({N_{RF}^3}T^3)$ & $\mathcal{O}\Big({N_r^2}(N_{RF}T-L)\Big)$ & $\mathcal{O}\Big((N_r^2+N_x^3){n_\phi}+N_r^2{n_\theta}L\Big)$\\
			HDS-IMRD-MUSIC & $\mathcal{O}({N_{RF}^2}{T^2}N_a)$ & $\mathcal{O}({N_{RF}^3}T^3)$ & $\mathcal{O}\Big({N_r^2}(N_{RF}T-L)\Big)$ & $\mathcal{O}\Big((N_r^2+N_x^3){{\bar n}_\phi}L\Big)$
			
			\\ \bottomrule
		\end{tabular}
		\label{table:sim_params}  
	\end{table*}  
 The adopted LS solution for azimuth DOA estimation can not only make full use of phase elements of $\mathbf{\hat a}_x(v)$, reducing the impact of estimated error effectively, but also can avoid the 1-D spectral search, further reducing the computational complexity. 

 \section{Computational Complexity and CRLB Analysis}   
	In this section, we first analyze the computational complexity of the proposed algorithms. We then derive the CRLB for 2-D DOA estimation with the adopted HDS structure. Finally, several   insights on DOA estimation under various configurations of HDS structure are provided, facilitating its practical design and applications.
 \subsection{Computational Complexity}  
	The computational complexity of the subspace-based DOA estimation algorithm mainly comes from SCM calculation  as well as  performing   EVD and spatial spectral search. For the HAD architecture, the size of the array covariance matrix is significantly reduced compared to the FD architecture, providing substantial advantages in both hardware implementation and computation requirements.
	
Specifically, for an $N_r=N_x\times N_z$ FD based UPA, the calculation of whole array SCM and its corresponding EVD with $TN_a$ total pilot symbol length require $\mathcal{O}(N_r^2TN_a+N_r^3)$ operations, and the 2-D spectral search required by the 2-D MUSIC algorithm has complexity $\mathcal{O}(N_r^2(N_r-L)+N_r^2{n_\theta}{n_\phi})$, where ${n_\theta}$ and ${n_\phi}$ represent the numbers of search grid points along azimuth and elevation angles. For the same sized HDS-based UPA shown in Fig. 1, the reconstructed observation ${\bf \tilde{Y}}\in \mathbb{C}^{N_{RF}T\times N_a}$ is used for DOA estimation. If we still apply the 2-D MUSIC algorithm for 2-D DOA estimation, the required complexity is roughly about $\mathcal{O}({N_{RF}^2}{T^2}N_a+{N_{RF}^3}T^3+{N_r^2}(N_{RF}T-L)+{N_r^2}{n_\theta}{n_\phi})$. For the adopted RD-MUSIC algorithm, it only requires 1-D spectral search, thus its total complexity is  $\mathcal{O}({N_{RF}^2}{T^2}N_a+{N_{RF}^3}T^3+{N_r^2}(N_{RF}T-L)+(N_r^2+N_x^3){n_\phi}+N_r^2{n_\theta}L)$. Finally, for the improved RD-MUSIC algorithm, since we exploit the closed-form solution for initial 1-D DOA estimation, small search spaced based 1-D spectral search and the LS solution for enhanced 2-D DOA estimation independently,  the major required complexity is reduced to $\mathcal{O}({N_{RF}^2}{T^2}N_a+{N_{RF}^3}T^3+{N_r^2}(N_{RF}T-L)+({N_{r}^2}+N_x^3){{\bar n}_\phi}L)$, where ${{\bar n}_\phi}\ll {{n}_\phi}$. A detailed complexity comparison is provided in Table \ref{table:sim_params}. It should be emphasized here that $N_{RF}T$ is normally set to be far less than $N_r$ so as to reduce system overhead, which is the key task for HAD architectures. As a result, it is concluded that the IMRD-MUSIC algorithm is computationally more efficient than the other algorithms.
 	\subsection{Cram\'{e}r-Rao Lower Bound}
To facilitate the derivation of the CRLB for 2-D DOA estimation, we first introduce the following notations. Let $\pmb{\theta}=\left[ {\theta}_1,\ldots,{\theta}_L\right]^T$ and $\pmb{\phi}=\left[ {\phi}_1,\ldots,{\phi}_L\right]^T$ denote the vectors of azimuth and elevation angles, respectively, and define the combined angle vector as $\pmb{\Theta}=\left[ \pmb{\theta}^T,\pmb{\phi}^T\right]^T$. For analytical convenience, we use $\pmb{\Phi}$ to represent the augmented analog combining matrix $\mathbf{\tilde{W}}^H$, and define its correlation matrix as $\mathbf{Q} = \pmb{\Phi}\pmb{\Phi}^H\in \mathbb{C}^{N_{RF} T \times N_{RF} T}$. The noise variance after pilot correction is denoted as $\sigma^{2}$. These definitions enable us to analyze how the analog combining matrix $\mathbf{W}_\mathrm{A}$ in \eqref{1} affects the CRLB.
	
Inspired by the CRLB analysis for 1-D DOAs in \cite{r39},   we extend its theoretical framework to 2-D DOA estimation with the proposed HDS architecture, which leads to several unique characteristics. First, our CRLB matrix is a $2L\times 2L$ matrix that captures both azimuth and elevation angles, in contrast to the $L\times L$ matrix in conventional 1-D scenarios. Moreover, the HDS architecture introduces distinct spatial sampling patterns that fundamentally affect the CRLB derivation. The  CRLB matrix for 2-D DOAs is derived as
	\begin{equation}\label{42}
		{\mathop{\rm CRLB}\nolimits}_{\mathrm{HDS}} = \frac{\sigma^2}{2} \left[ \mathcal{R}\left\lbrace \sum\limits_{n = 1}^{{N_a}} \mathbf{\tilde{S}}^{H}(n)\mathbf{\tilde{B}}^{H} \pmb{\Pi}_{\mathbf{\tilde{E}}} \mathbf{\tilde{B}}\mathbf{\tilde{S}}(n)\right\rbrace \right]^{-1},
	\end{equation}
	where
	\begin{equation}\label{43}
		\mathbf{\tilde{B}} = \mathbf{Q}^{-\frac{1}{2}}\pmb{\Phi}\mathbf{B},
	\end{equation}
	\begin{equation}\label{44}
		\mathbf{B}= \left[ \frac{\partial \mathbf{a}}{\partial \theta} \bigg|_{\theta = \theta_1}, \dots, \frac{\partial \mathbf{a}}{\partial \theta} \bigg|_{\theta = \theta_L},\frac{\partial \mathbf{a}}{\partial \phi} \bigg|_{\phi = \phi_1},\dots, \frac{\partial \mathbf{a}}{\partial \phi} \bigg|_{\phi = \phi_L}\right],		
	\end{equation}
	\begin{equation}\label{45}
		\pmb{\Pi}_{\mathbf{\tilde{E}}} = \mathbf{I}_{N_{RF}T} - \mathbf{\tilde{A}} \left( \mathbf{\tilde{A}}^H \mathbf{\tilde{A}} \right)^{-1} \mathbf{\tilde{A}}^H,
	\end{equation}
	\begin{equation}\label{46}
		\mathbf{\tilde{A}}=\mathbf{Q}^{-\frac{1}{2}}\pmb{\Phi}\mathbf{A}_r,
	\end{equation}
	\begin{equation}\label{47}
		\mathbf{\tilde{S}}(n) = \mathrm{diag}\{s_1(n),\cdots,s_L(n),s_1(n),\cdots,s_L(n)\},
	\end{equation}
	with $\mathbf{a}=\mathbf{a}(\theta, \phi)$ representing the steering vector as defined in (6), $\mathbf{\tilde{S}}(n)\in\mathbb{C}^{ 2L\times 2L}$ the  diagonal matrix formed by the $n$-th column of   $\mathbf{S}$, and $s_\ell(n)$ the element in the $\ell$-th row and $n$-th column of   $\mathbf{S}$.
	
	For subsequent analysis as well as comparison, the CRLB for FD architecture is also provided. Different from HDS architectures, the signal reconstruction of observations is not necessary. Assuming that the fully digital architecture employs $T_d$ pilot signals. The equivalent transmitted signal matrix $\mathbf{S}$ is transformed into $\mathbf{\bar S} \in \mathbb{C}^{ L\times N_a T_d}$, where $\mathbf{\bar S}$ represents $\mathbf{S}$ repeated $T_d$ times along the time dimension. Then, the corresponding CRLB matrix for DOA estimation is 
	\begin{equation}\label{48}
		{\mathop{\rm CRLB}\nolimits}_{\mathrm{FD}}=\frac{\sigma^2}{2}\left[\mathcal{R}\Big\{\sum_{n=1}^{N_a T_d}\mathbf{\bar S}^H(n)\mathbf{B}^H\mathbf{P}_\mathbf{A}^\perp\mathbf{B}\mathbf{\bar S}(n)\Big\}\right]^{-1},
	\end{equation}
	where
	\begin{equation}\label{49}
		\mathbf{P}_\mathbf{A}^\perp = \mathbf{I}_{N_r} - \mathbf{A}_r \left( \mathbf{A}_r^H \mathbf{A}_r \right)^{-1} \mathbf{A}_r^H,
	\end{equation}
	\begin{equation}\label{50}
		\mathbf{\bar S}(n) = \mathrm{diag}\{\bar s_1(n),\cdots,\bar s_L(n),\bar s_1(n),\cdots,\bar s_L(n)\},
	\end{equation}
	with $\mathbf{\bar S}(n)\in\mathbb{C}^{2L\times 2L}$ being the diagonal matrix formed by the $n$-th column of   $\mathbf{\bar S}$, and $\bar s_\ell(n)$ the element in the $\ell$-th row and $n$-th column of   $\mathbf{\bar S}$.
	\subsection{Performance Analysis and Comparison for Different Architectures/Connections via CRLB}
	Based on the derived 2-D DOA CRLB, we present two fundamental theoretical findings that provide novel insights into the performance of different array architectures in THz UM-MIMO systems.
	  
	To simplify the analysis, the following assumptions with a single source scenario are considered: 1) the analog combining matrix  $\mathbf{W}_\mathrm{A}$ in \eqref{1} remains unchanged during each pilot duration but varies across different pilots; 2) the total number of switches used across $T$ different analog combining matrices remains constant; 3) each RF chain probabilistically connects to subarrays with equal likelihood, and different RF chains are connected to the same number of subarrays. Under such assumptions, each row of $\pmb{\Phi}$ contains the same number of non-zero elements, with their positions randomly appearing in block patterns, and $\mathbf{Q}$ is approximated as $\mathbf{Q}\approx\rho\mathbf{I}$ \footnote{Because of the random phase shift characteristics of the adopted HDS architectures, and the fact that the cross-correlation value of the random phase process is much lower than its autocorrelation value, it is important to clarify that such an approximation is reasonable.}.
	
	\emph{Theorem 2:} For the HDS architectures with random phase shifts, the following system designs exhibit almost the same estimation performance: 1) HFC architecture; 2) HS architecture; 3) HDS architecture with overlapping subarray design; 4) HDS architecture with randomly selected switches, where the overlapping subarray design in 3) means that the connection between the first RF chain and several subarrays is randomly determined, while the connections of the other RF chains are shifted by a fixed unit. If the subarray index exceeds the total number of subarrays, a modulo operation is used to determine the position of the connected subarray.
	\begin{proof}
		See Appendix A.
	\end{proof}
	This theorem reveals that the optimal estimation performance can be achieved with reduced hardware complexity through strategic switch deployment, providing new guidelines for flexible HDS architecture design in THz systems.

	\emph{Theorem 3:} The CRLB for HDS architecture and that for FD architecture with an equivalent number of antennas satisfy
	\begin{equation}\label{51}
		{\mathop{\rm CRLB}\nolimits}_{\mathrm{HDS}}\approx\frac{N_r T_d}{N_\mathrm{RF}T}{\mathop{\rm CRLB}\nolimits}_{\mathrm{FD}}.
	\end{equation}
	\begin{proof}
		See Appendix B.
	\end{proof}
While this relationship shares a similar form with the 1-D case in \cite{r39}, our analysis extends it to 2-D scenarios with several key distinctions: \textit{1) the relationship now encompasses both azimuth and elevation angles, introducing new spatial dimensions not considered in previous work; 2) the impact of HDS architecture on 2-D estimation reveals unique trade-offs between hardware complexity and performance; 3) the ratio provides new insights for optimizing system parameters in THz UM-MIMO applications, particularly in scenarios where both angular dimensions are crucial. These findings offer practical guidance for system designers in balancing antenna configuration, RF chain allocation, and pilot signal design in 2-D DOA estimation.}
	\begin{figure}[!t]
		\vspace{-2mm}\centering
		\includegraphics[width=3.35in]{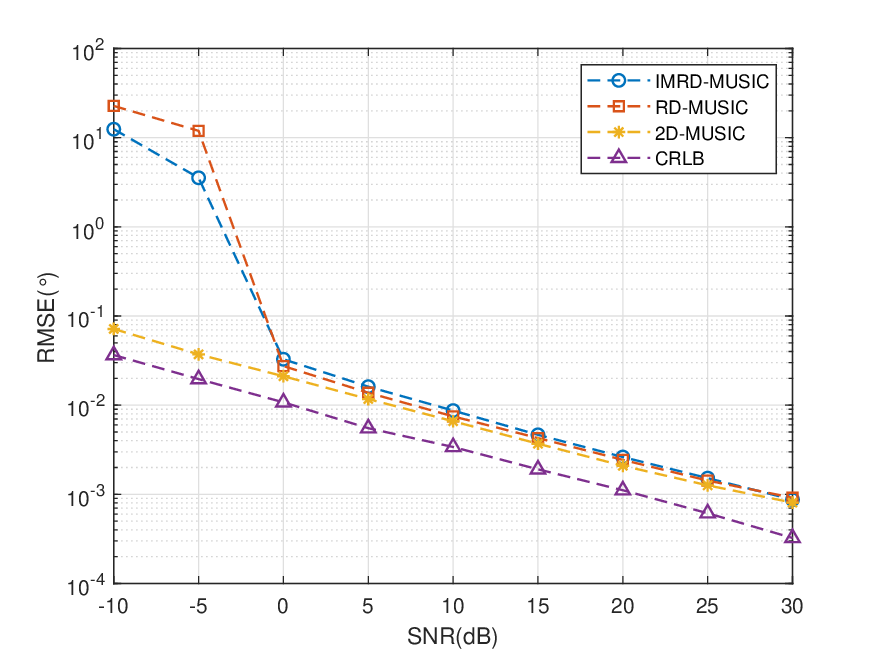}
		\caption{RMSE of DOA estimates versus SNR, with $N_\mathrm{RF}=8$, $T=12$.}
		\label{Fig.2.}  \vspace{-2mm}
	\end{figure}
	\begin{figure}[!t]
		\vspace{-2mm}	\centering
		\includegraphics[width=3.35in]{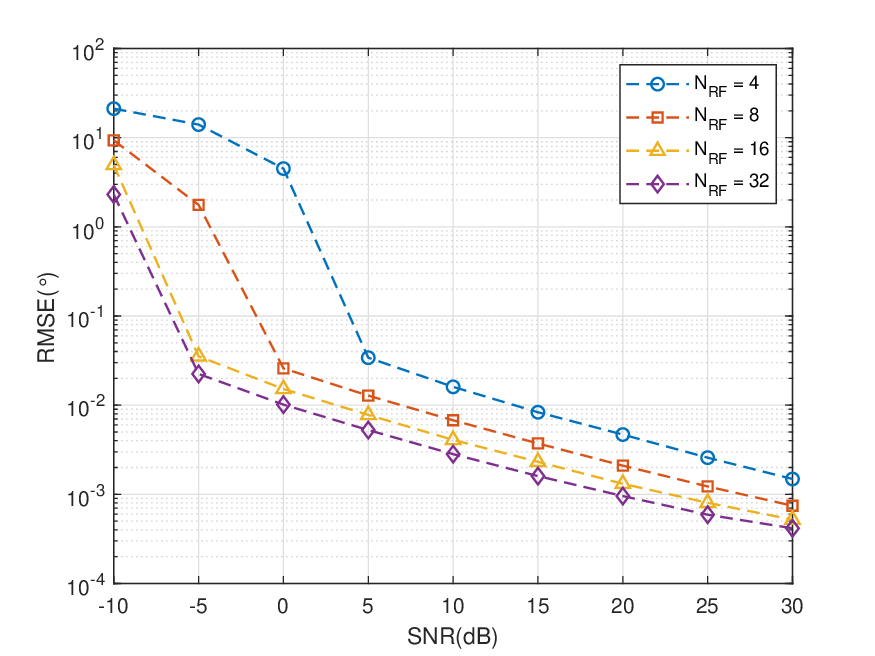}
		\caption{RMSE of DOA estimates versus SNR and $N_\mathrm{RF}$, with $T=16$.}
		\label{Fig.3.}  \vspace{-2mm}
	\end{figure}
	\vspace{-1mm} \section{Simulation Results}
	In this section, the DOA estimation performance of the proposed algorithm is assessed. Unless otherwise specified, the simulation configurations are set as follows: the central frequency is $f$=1.0 THz, the total number of array antennas is $N_r=1024$ with $N_x=N_z=32$, the number of RF chains $N_\mathrm{RF}$ varies in 4, 8, 16 and 32, whereas the number of pilot signals $T$ varies in 8, 12, 16 and 20. In addition, three propagation paths composed of one LOS path and two NLOS paths are considered. The channel attenuation for the NLOS paths is randomly generated between 5 dB and 10 dB. The azimuth and elevation angles are randomly generated, following uniform distributions over $[-180^\circ, 180^\circ]$ and $[-90^\circ, 90^\circ]$, respectively. The root mean square error (RMSE) based on 1000 Monte-Carlo trials is used as the DOA estimation performance metric. In the simulations, we mainly vary the number of RF chains and pilot signals to observe their impacts on estimation accuracy. To ensure broad applicability, the switch states are random. Meanwhile, the number of closed switches is set to be half of the total number of switches, simulating a $50\%$ probability of each switch being randomly closed. Additionally, we ensure that every RF chain and subarray are utilized to avoid hardware waste.

	The first simulation evaluates DOA estimation performance of different algorithms versus signal-to-noise ratio (SNR) with $N_\mathrm{RF}=8$ and $T=12$, and the results are plotted in Fig. 2. It is observed that the RMSEs of all algorithms decrease with the increase of SNR. Moreover, the   RD-MUSIC and IMRD-MUSIC achieve nearly identical estimation performance when SNR $\geq$ 0 dB and follow HDS based 2D-MUSIC algorithm very closely, demonstrating the effectiveness of the two proposed solutions. On the other hand, IMRD-MUSIC outperforms RD-MUSIC in low SNRs, which is attributed to the special design of the adopted HDS architecture and the application of the LS solution in (40). Combining with the computational efficiency of   IMRD-MUSIC, we can regard IMRD-MUSIC as a practical solution well-suited for real-world applications.
	\begin{figure}[!t]
		\centering
		\includegraphics[width=3.35in]{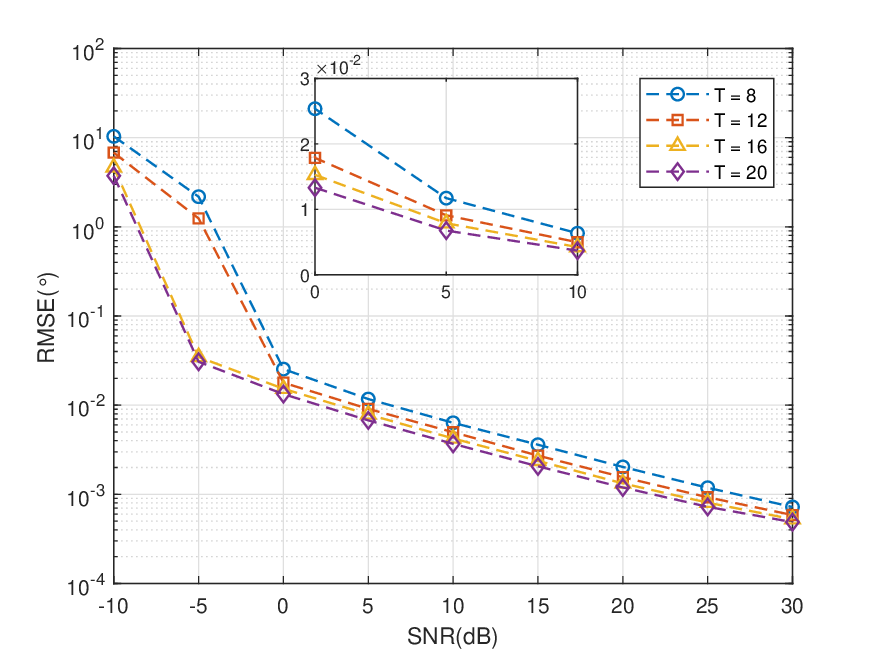}
		\caption{RMSE of DOA estimates versus SNR and $T$, with $N_\mathrm{RF}=16$.}
		\label{Fig4.}  \vspace{-2mm}
	\end{figure} \begin{figure}[!t]
		\vspace{-3mm} 	\centering
		\includegraphics[width=3.35in]{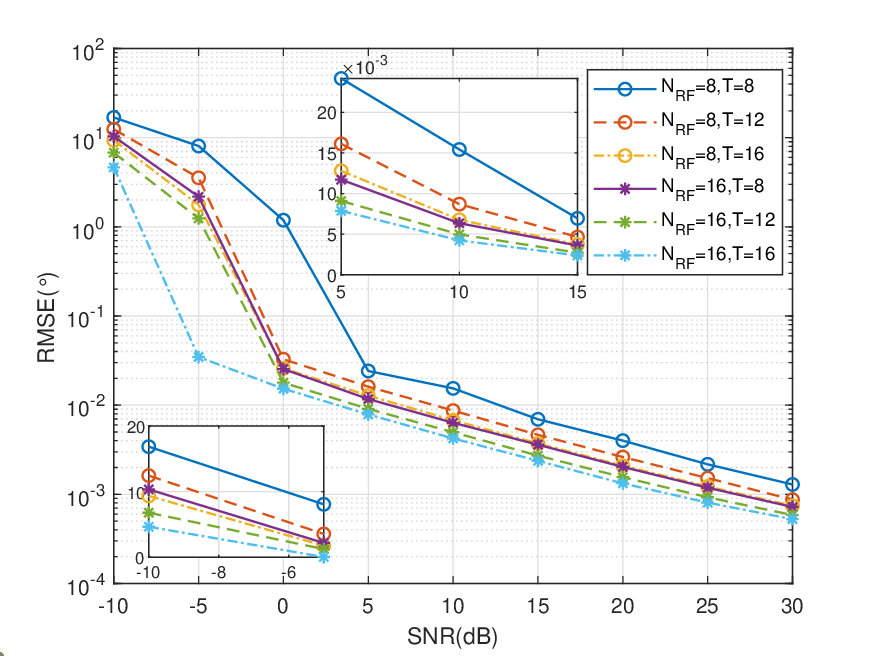}
		\caption{RMSE of DOA estimates versus SNR, $N_\mathrm{RF}$ and $T$.}
		\label{Fig.5.} 
	\end{figure}	
	
	The second simulation assesses the impact of the number of RF chains $N_\mathrm{RF}$ and the number of pilot signals $T$ on the performance of   IMRD-MUSIC separately, whose RMSE curves are shown in Figs. 3 and 4, respectively, from which we can see that the DOA estimation accuracy increases greatly with the increase of $N_\mathrm{RF}$ and $T$ at the beginning; however, with the continuous increase of both $N_\mathrm{RF}$ and $T$, the degree of improvement diminishes progressively.
	
	The third simulation further evaluates the impact of both $N_\mathrm{RF}$ and $T$ on DOA estimation performance. It is seen from Fig. 5 that the performance of the proposed solution with $N_\mathrm{RF}=16$ indeed achieves higher accuracy than that with $N_\mathrm{RF}=8$. Notably, the   performance is very similar when $N_\mathrm{RF}=8, T=16$ and $N_\mathrm{RF}=16, T=8$. This suggests that when $N_\mathrm{RF}$ is insufficient, increasing $T$ can partially offset the accuracy loss caused by hardware limitations. Nevertheless, it is important to emphasize that the simply increasing $N_\mathrm{RF}$ and/or $T$ is not the optimal choice, since the performance improvement could be limited, but the system overhead increases significantly. As a result, it is crucial to carefully balance the trade-off between RF chain and pilot overhead  in practice.
	\begin{figure}[!t]
		\vspace{-2mm}	\centering
		\includegraphics[width=3.35in]{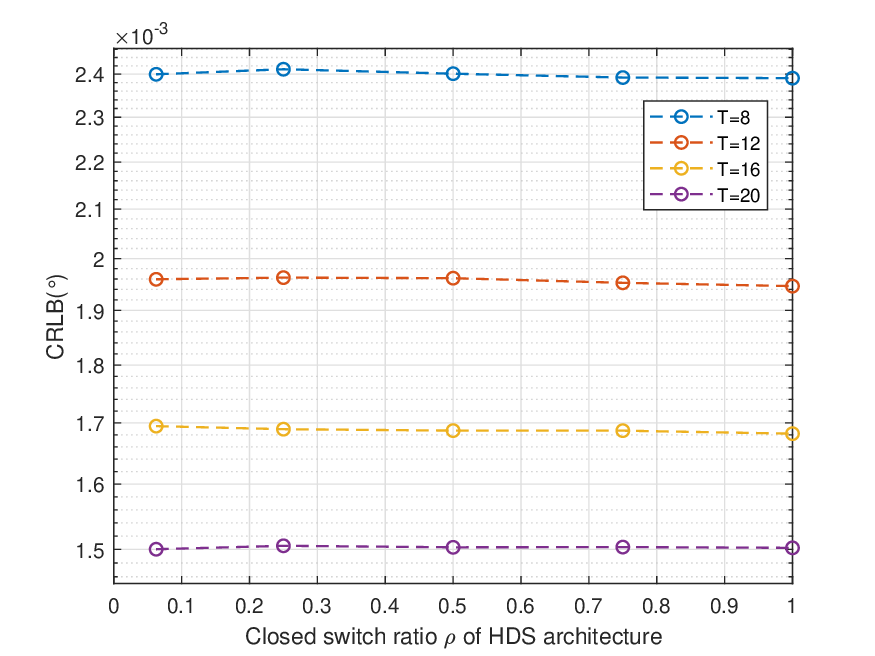}
		\caption{CRLB under different numbers of closed switches.}
		\label{Fig.6.}
	\end{figure}
	
	The fourth simulation tests the performance across different architectures through CRLB comparison, at $N_\mathrm{RF}=16$. The corresponding result is shown in Fig. 6, where $\rho$ increases from $1/N_\mathrm{RF}$ to 1, i.e., the array architecture changes from HS connection to HFC connection. It is seen that the CRLB nearly remains constant as the number of switches increases for arbitrary $T$, which is consistent with  Theorem 2.
	
	The last simulation studies the DOA estimation performance of the   IMRD-MUSIC   various number of antennas and $T$, where SNR and $N_\mathrm{RF}$ are fixed at 10 dB and 16, respectively. The results are plotted in Fig. 7 and we see that,  as the number of antennas increases, the RMSE  decreases significantly, highlighting the dominant effect of array aperture on estimation accuracy. However, it is noteworthy that when $T=8$, the RMSE for 1280 antennas is actually higher than that for 1024 antennas. This is because that the number of RF chains remains unchanged, meaning the dimension of the observed signal stays constant, but the increase in the number of antennas introduces phase ambiguity, thereby affecting the final estimation accuracy. Under such a circumstance, increasing the number of pilots used in the estimation process can help mitigate this effect to certain extent.
	\vspace{-3mm} \begin{figure}[!t]
		\vspace{-5mm} \centering
		\includegraphics[width=3.3in]{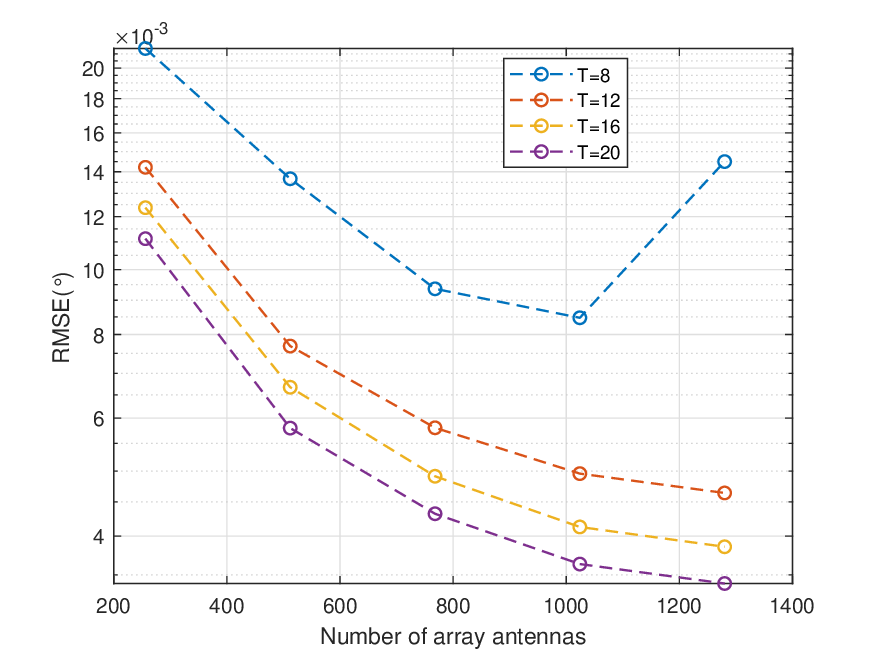}
		\caption{RMSE of DOA estimates versus array antenna number and $T$ at
			SNR=10 dB. }
		\label{Fig.7.}  \vspace{-3mm}
	\end{figure}
	\section{Conclusion}
	In this article, an HDS architecture with great flexibility has been designed, providing a solid hardware foundation for achieving high-efficiency and high-accuracy DOA estimation within the framework of the HAD architecture. With the aid of  the designed HDS architecture, the phase compensation based RD-MUSIC as well as improved RD-MUSIC utilizing the special structure of HDS architecture and sparse nature of transmission channels have been proposed, yielding a high accuracy DOA estimation with reduced complexity. The corresponding CRLB is derived and performance analysis for different architectures/connections via CRLB are also conducted, providing important theoretical insights for flexibly configuring the HAD architecture based on practical situations. Simulation results configured with various array configurations have been provided to demonstrate the superiority and effectiveness of the developed solutions. We acknowledge the assistance of Anthropic's Claude 3.5 Sonnet  in refining portions of the manuscript text.
	\section*{Appendix A \\ Proof of Theorem 2}
	For the HDS architectures, we have  $\mathbf{\tilde{A}}=\mathbf{\tilde{a}}$, $\mathbf{\tilde{a}}=\mathbf{Q}^{-\frac{1}{2}}\pmb{\Phi}\mathbf{a}$, $\mathbf{\tilde{B}}=[ \mathbf{\tilde{b}}_1,\mathbf{\tilde{b}}_2] $, $\mathbf{\tilde{b}}_1=\mathbf{Q}^{-\frac{1}{2}}\pmb{\Phi}\mathbf{b}_1$, $\mathbf{\tilde{b}}_2=\mathbf{Q}^{-\frac{1}{2}}\pmb{\Phi}\mathbf{b}_2$, $\mathbf{b}_1=\frac{\partial \mathbf{a}}{\partial \theta}= j\frac{2\pi}{\lambda}  \cos\theta\cos\phi \mathbf{D}_1 \mathbf{a}$, $\mathbf{b}_2=j\frac{2\pi}{\lambda} \mathbf{D}_2 \mathbf{a}$, and
	\begin{equation}\label{52}
		\mathbf{D}_1 = \mathrm{diag}\{d_1,\cdots,d_{N_{x}},\cdots,d_1,\cdots,d_{N_{x}}\},
	\end{equation}
	\begin{multline}\label{53}
		\mathbf{D}_2 = \mathrm{diag}\{d'_1 \cos \phi-d_1\sin\theta\sin\phi,\cdots,d'_1 \cos \phi- \\
		d_{N_{x}}\sin\theta\sin\phi,\cdots,d'_{N_{z}}\cos \phi-d_1\sin\theta\sin\phi \cos \phi,\\
		\cdots,d'_{N_{z}}\cos \phi-d_{N_{x}}\sin\theta\sin\phi\},
	\end{multline}
	with $d_{n_{x}}=\left( n_{x}-1\right)d $ and $d'_{n_{z}}=\left( n_{z}-1\right)d $.
	
	The Fisher information matrix corresponding to \eqref{42} is rewritten as
	\begin{equation}\label{54}
		{\mathop{\rm CRLB}\nolimits}_{\mathrm{HDS}}^{-1} = \mathcal{R}\left\lbrace\begin{bmatrix}
			\mathbf{F}_{\pmb{\theta}\pmb{\theta},\mathrm{HDS}} & \mathbf{F}_{\pmb{\theta}\pmb{\phi},\mathrm{HDS}} \\
			\mathbf{F}_{\pmb{\phi}\pmb{\theta},\mathrm{HDS}} & \mathbf{F}_{\pmb{\phi}\pmb{\phi},\mathrm{HDS}},
		\end{bmatrix}\right\rbrace
	\end{equation}
	where
	\begin{equation}\label{55}
		\mathbf{F}_{\pmb{\theta}\pmb{\theta},\mathrm{HDS}} = \frac{2N_a \hat{p} \left( \|\tilde{\mathbf{a}}\|_2^2 \|\tilde{\mathbf{b}}_1\|_2^2 - \left| \tilde{\mathbf{b}}_1^H \tilde{\mathbf{a}} \right|^2 \right)}{\sigma^2 \|\tilde{\mathbf{a}}\|_2^2},
	\end{equation}
	\begin{equation}\label{56}
		\mathbf{F}_{\pmb{\phi}\pmb{\phi},\mathrm{HDS}} = \frac{2N_a \hat{p} \left( \|\tilde{\mathbf{a}}\|_2^2 \|\tilde{\mathbf{b}}_2\|_2^2 - \left| \tilde{\mathbf{b}}_2^H \tilde{\mathbf{a}} \right|^2 \right)}{\sigma^2 \|\tilde{\mathbf{a}}\|_2^2},
	\end{equation}
	\begin{equation}\label{57}
		\mathbf{F}_{\pmb{\theta}\pmb{\phi},\mathrm{HDS}} = \frac{2N_a \hat{p} \left( \|\tilde{\mathbf{a}}\|_2^2 \tilde{\mathbf{b}}_1^H \tilde{\mathbf{b}}_2-\tilde{\mathbf{b}}_1^H\tilde{\mathbf{a}}\tilde{\mathbf{a}}^H\tilde{\mathbf{b}}_2 \right)}{\sigma^2 \|\tilde{\mathbf{a}}\|_2^2},
	\end{equation}
	\begin{equation}\label{58}
		\mathbf{F}_{\pmb{\phi}\pmb{\theta},\mathrm{HDS}} = \frac{2N_a \hat{p} \left( \|\tilde{\mathbf{a}}\|_2^2 \tilde{\mathbf{b}}_2^H \tilde{\mathbf{b}}_1-\tilde{\mathbf{b}}_2^H\tilde{\mathbf{a}}\tilde{\mathbf{a}}^H\tilde{\mathbf{b}}_1 \right)}{\sigma^2 \|\tilde{\mathbf{a}}\|_2^2},
	\end{equation}
	with $\hat{p} = \frac{1}{N_a } \sum_{n=1}^{N_a } | s(n)|^2$ being the estimate of $p = \mathbb{E} \{ |s(n)|^2 \}$.
	
	For $\mathbf{F}_{\pmb{\theta}\pmb{\theta},\mathrm{HDS}}$, we calculate $\|\tilde{\mathbf{a}}\|_2^2$,  $\|\tilde{\mathbf{b}}_1\|_2^2$ and $\left| \tilde{\mathbf{b}}_1^H \tilde{\mathbf{a}} \right|^2$ separately. First, 
	$\|\tilde{\mathbf{a}}\|_2^2$ is computed by
	\begin{equation}\label{59}
		\begin{gathered}
			\left\|\tilde{\mathbf{a}}\right\|_{2}^{2}= \frac{1}{\rho} \mathbf{a}^H\mathbf{\Phi}^H \mathbf{\Phi}\mathbf{a}.
		\end{gathered}
	\end{equation}
	
	Let $\Omega(k)$ denote the set of positions of non-zero elements in the $k$-th row of matrix $\mathbf{\Phi}$. Then, we can express the $k$-th element of $\mathbf{\Phi}\mathbf{a}$ as $ \frac{1}{\sqrt{N_r}} \sum_{n \in \Omega(k)} e^{j\left(w_{n}-\alpha_{k,n}\right)} $,  which yields
	\begin{equation}\label{60}
		\begin{aligned}
			&\left\|\tilde{\mathbf{a}}\right\|_{2}^{2}= \frac{1}{\rho N_r}\sum_{k=1}^{N_\mathrm{RF} T}  \Big(\sum_{n \in \Omega(k)}e^{j\left(\alpha_{k,n}-w_n\right)}\sum_{n^{\prime}\in \Omega(k)}e^{j\left(w_{n'}-\alpha_{k,n^{\prime}}\right)}\Big) \\
			&=N_\mathrm{RF} T+\frac{1}{ N_r}\sum_{k=1}^{N_\mathrm{RF} T}\sum_{n \in \Omega(k)}\sum_{n^{\prime}\neq n}e^{j(\alpha_{k,n}-w_n)+j(w_{n'}-\alpha_{k,n^{\prime}})}.
		\end{aligned}
	\end{equation}
	The second term is zero due to property that the characteristic function of $e^{j\alpha}$ with $\alpha\sim\mathcal{U}(0, 2\pi)$ is zero \cite{r40}.
	
	The term $\|\tilde{\mathbf{b}}_1\|_2^2$ is:
	\begin{equation}\label{61}
		\left\|\mathbf{\tilde{b}}_1\right\|_2^2=\frac{4\pi^2\cos^2\theta\cos^2\phi}{\rho \lambda^2}\mathbf{a}^H\mathbf{D}_1^H\mathbf{\Phi}^H \mathbf{\Phi}\mathbf{D}_1\mathbf{a},
	\end{equation}
	where $\mathbf{a}^{H}\mathbf{D}_1^{H}\mathbf{\Phi}^{H}\mathbf{\Phi}\mathbf{D}_1\mathbf{a}=\|\mathbf{\Phi}\mathbf{D}_1\mathbf{a}\|_2^2$ is expanded as
	\begin{equation}\label{62}
		\begin{aligned}
			&\|\mathbf{\Phi}\mathbf{D}_1\mathbf{a}\|_{2}^{2}= \\  
			&=\frac{1}{N_r}\sum_{k=1}^{N_\mathrm{RF} T}\sum_{n\in \Omega(k)}\bar{d}^2_{n}+\\
			&\frac{1}{ N_r}\sum_{k=1}^{N_\mathrm{RF} T}\sum_{n \in \Omega(k)}\sum_{n^{\prime}\neq n}e^{j(\alpha_{k,n}-w_n)+j(w_{n'}-\alpha_{k,n^{\prime}})},
		\end{aligned}
	\end{equation}
	the $k$-th element of $\mathbf{\Phi}\mathbf{D}_1\mathbf{a}$  is $ \frac{1}{\sqrt{N_r}} \sum_{n \in \Omega(k)} \bar{d}_n e^{j\left(w_n-\alpha_{k,n}\right)}$, with $\bar{d}_n$ representing the $n$-th diagonal element of $\mathbf{D}_1$. Similarly, by applying the properties of the characteristic function, the second term is eliminated, and the first term is approximated by $\frac{N_\mathrm{RF} T}{N_r} \rho \sum_{n=1}^{N_r}\bar{d}^2_n$. Note that each RF chain randomly selects the same number of subarrays to connect across $N_\mathrm{RF} T$ observation dimensions. Subsequently, it is derived that
	\begin{equation}\label{63}
		\left\|\mathbf{\tilde{b}}_1\right\|_2^2\approx \frac{4\pi^2\cos^2\theta\cos^2\phi N_\mathrm{RF} T}{N_r \lambda^2}\sum_{n=1}^{N_r}\bar{d}^2_n,
	\end{equation}
	and $\left| \tilde{\mathbf{b}}_1^H \tilde{\mathbf{a}} \right|^2$ is given by
	\begin{equation}\label{64}
		\left| \tilde{\mathbf{b}}_1^H \tilde{\mathbf{a}} \right|^2=\frac{4\pi^2\cos^2\theta\cos^2\phi}{\rho^2\lambda^2}\Big|\mathbf{a}^H\mathbf{D}_1^H\mathbf{\Phi}^\mathrm{H}\mathbf{\Phi}\mathbf{a}\Big|^2,
	\end{equation}
	with $\Big|\mathbf{a}^H\mathbf{D}_1^H\mathbf{\Phi}^H\mathbf{\Phi}\mathbf{a}\Big|^2$ being calculated as:
	\begin{equation}\label{65}
		\begin{aligned}			&\left|\mathbf{a}^H\mathbf{D}_1^H\mathbf{\Phi}^H\mathbf{\Phi}\mathbf{a}\right|^{2}=\frac{1}{N_r^{2}}\Bigg|\sum_{k=1}^{N_\mathrm{RF} T}  \Big(\sum_{n \in \Omega(k)}\bar{d}_{n}e^{j\left(\alpha_{k,n}-w_n\right)}\\
			&\sum_{n^{\prime}\in \Omega(k)}e^{j\left(w_{n'}-\alpha_{k,n^{\prime}}\right)}\Big)\Bigg|^{2}\approx\frac{1}{N_r^{2}}\Bigg|\rho N_\mathrm{RF}T\sum_{n=1}^{N_r}\bar{d}_{n}\Bigg|^{2}.
		\end{aligned}
	\end{equation}
	
	Substituting  \eqref{60},  \eqref{61} and   \eqref{64} nto   \eqref{55}, and defining $K=N_\mathrm{RF}/N_r$, we have
	\begin{equation}\label{66}
		\mathbf{F}_{\pmb{\theta}\pmb{\theta},\mathrm{HDS}} \approx K \frac{8N_a T\hat{p}\pi^2\cos^2\theta\cos^2\phi}{ N_r\sigma^2\lambda^2}\Big(N_r\sum_{n=1}^{N_r}{\bar{d}}_{n}^{2}-\Big(\sum_{n=1}^{N_r}{\bar{d}}_{n} \Big)^2\Big).
	\end{equation}
	
	Similarly, $\mathbf{F}_{\pmb{\phi}\pmb{\phi},\mathrm{HDS}}$ is approximately expressed as
	\begin{equation}\label{67}
		\mathbf{F}_{\pmb{\phi}\pmb{\phi},\mathrm{HDS}} \approx K\frac{8N_a T\hat{p}\pi^2}{ N_r\sigma^2\lambda^2}\Big(N_r\sum_{n=1}^{N_r}{\tilde{d}}_{n}^{2}-\Big(\sum_{n=1}^{N_r}{\tilde{d}}_{n} \Big)^2\Big), \vspace{-2mm}
	\end{equation}
	where $\tilde{d}_n$ represents the $n$-th diagonal element of $\mathbf{D}_2$.
	
	For $\mathbf{F}_{\pmb{\theta}\pmb{\phi},\mathrm{HDS}}$ and $\mathbf{F}_{\pmb{\phi}\pmb{\theta},\mathrm{HDS}}$, it is further derived that $\tilde{\mathbf{b}}_1^H \tilde{\mathbf{b}}_2$ is expressed as
	\begin{equation}\label{68}
		\tilde{\mathbf{b}}_1^H \tilde{\mathbf{b}}_2 = \frac{4\pi^2\cos\theta\cos\phi}{\rho\lambda^2}\mathbf{a}^H\mathbf{D}_1^H\mathbf{\Phi}^H\mathbf{\Phi}\mathbf{D}_2\mathbf{a},
	\end{equation}
	with $\mathbf{a}^H\mathbf{D}_1^H\mathbf{\Phi}^H\mathbf{\Phi}\mathbf{D}_2\mathbf{a}$ computed by
	\begin{equation}\label{69}
		\begin{aligned}
			&\mathbf{a}^H\mathbf{D}_1^H\mathbf{\Phi}^H\mathbf{\Phi}\mathbf{D}_2\mathbf{a}  \\
			&= \frac{1}{N_r} \sum_{k=1}^{N_\mathrm{RF} T}\sum_{n\in \Omega(k)} \bar{d}_n \tilde{d}_n
			+ \\
			&\frac{1}{N_r} \sum_{k=1}^{N_\mathrm{RF} T} \sum_{n \in \Omega(k)} \sum_{\substack{ n^{\prime} \neq n}} \bar{d}_n \tilde{d}_{n'}
			e^{j(\alpha_{k,n} - w_n) + j(w_{n'} - \alpha_{k,n'})} \\
			&\approx\frac{\rho N_\mathrm{RF}T}{N_r} \sum_{n=1}^{N_r} \bar{d}_n \tilde{d}_n,
		\end{aligned}
	\end{equation}

	Next, the term $\tilde{\mathbf{b}}_1^{H}\tilde{\mathbf{a}}\tilde{\mathbf{a}}^{H}\tilde{\mathbf{b}}_2$ is expressed as
	\begin{equation}\label{72}
		\tilde{\mathbf{b}}_1^{H}\tilde{\mathbf{a}}\tilde{\mathbf{a}}^{H}\tilde{\mathbf{b}}_2 = \frac{4\pi^2\cos\theta\cos\phi}{\rho^2\lambda^2}\mathbf{a}^{H}\mathbf{D}_1^{H}\mathbf{\Phi}^{H}\mathbf{\Phi}
		{\mathbf{a}}{\mathbf{a}}^{H}\mathbf{\Phi}^{H}\mathbf{\Phi}\mathbf{D}_2\mathbf{a}.
	\end{equation}
	
Employing \eqref{65}, $\mathbf{a}^{H}\mathbf{D}_1^{H}\mathbf{\Phi}^{H}\mathbf{\Phi}{\mathbf{a}}{\mathbf{a}}^{H}\mathbf{\Phi}^{H}\mathbf{\Phi}\mathbf{D}_2\mathbf{a}$ is approximately expressed as
	\begin{equation}\label{73}
		\begin{aligned}
			&\mathbf{a}^\mathrm{H}\mathbf{D}_1^\mathrm{H}\mathbf{\Phi}^\mathrm{H}\mathbf{\Phi}{\mathbf{a}}{\mathbf{a}}^\mathrm{H}\mathbf{\Phi}^\mathrm{H}\mathbf{\Phi}\mathbf{D}_2\mathbf{a}\approx\frac{(\rho N_\mathrm{RF}T)^2}{N_r^2}\Big(\sum_{n=1}^{N_r}\bar{d}_{n}\sum_{n=1}^{N_r}\tilde{d}_{n}\Big).
		\end{aligned}
	\end{equation}
	
	Accordingly, we have
	\begin{equation}\label{74}
		\begin{aligned}
			&\mathbf{F}_{\pmb{\theta}\pmb{\phi},\mathrm{HAD}} = \mathbf{F}_{\pmb{\phi}\pmb{\theta},\mathrm{HAD}} \\
			&\approx K\frac{8N_a T \hat{p}\pi^2\cos\theta\cos\phi}{N_r\sigma^2\lambda^2}\Big(N_r \sum_{n=1}^{N_r}{\bar{d}}_{n}{\tilde{d}}_{n}
			- \\ &\Big(\sum_{n=1}^{N_r}{\bar{d}}_{n}\Big)\Big(\sum_{n=1}^{N_r}{\tilde{d}}_{n}\Big)\Big).
		\end{aligned}
	\end{equation}
	
	By examining the explicit expressions of the CRLB terms, we observe that different architecture connections (with $\rho$ ranging from $1 / N_\mathrm{RF}$ to 1) under the given assumptions mainly affect the magnitudes of different elements in $\tilde{\mathbf{a}}$, $\tilde{\mathbf{b}}_1$ and $\tilde{\mathbf{b}}_2$, but this effect is eliminated in the final expression. By designing architecture connections that satisfy the stipulated conditions (i.e., the number of non-zero element in each column of $\mathbf{\Phi}$ is the same), the four considered architectures in Theorem 2 achieve almost the same performance.
	
	
	It is necessary to point out that each column of   $\mathbf{W}_{\mathrm{A}}$ for the first three connections has the same number of non-zero elements for each pilot signal. However, the HDS architecture with a random switch selection design breaks this limitation, as long as the total number of elements across the $T$ pilots meets the fundamental requirements (i.e., the number of non-zero elements in each column of $\mathbf{\Phi}$ is the same), its estimation performance is almost consistent with those of the first three connections, offering greater design flexibility. Moreover, by extending the observation dimension to $N_\mathrm{RF} T$, the statistical properties of the non-zero elements are optimized, making them closer to an ideal uniform distribution.
	\section*{Appendix B \\ Proof of Theorem 3}
	For FD architectures, we have $\mathbf{A}_r=\mathbf{a}$  and  $\mathbf{B}=\left[ \mathbf{b}_1,\mathbf{b}_2\right] $. The Fisher information matrix corresponding to  \eqref{41} is expressed as
	\begin{equation}\label{75}
		{\mathop{\rm CRLB}\nolimits}_{\mathrm{FD}}^{-1} = \mathcal{R}\left\lbrace \begin{bmatrix}
			\mathbf{F}_{\pmb{\theta}\pmb{\theta},\mathrm{FD}} & \mathbf{F}_{\pmb{\theta}\pmb{\phi},\mathrm{FD}} \\
			\mathbf{F}_{\pmb{\phi}\pmb{\theta},\mathrm{FD}} & \mathbf{F}_{\pmb{\phi}\pmb{\phi},\mathrm{FD}}
		\end{bmatrix}\right\rbrace
	\end{equation}
	where  
	\begin{equation}\label{76}
		\mathbf{F}_{\pmb{\theta}\pmb{\theta},\mathrm{FD}} = \frac{2N_a T_d \hat{p} \left( \|{\mathbf{a}}\|_2^2 \|{\mathbf{b}}_1\|_2^2 - \left| {\mathbf{b}}_1^H {\mathbf{a}} \right|^2 \right)}{\sigma^2 \|{\mathbf{a}}\|_2^2}
	\end{equation}
	\begin{equation}\label{77}
		\mathbf{F}_{\pmb{\phi}\pmb{\phi},\mathrm{FD}} = \frac{2N_a T_d\hat{p} \left( \|{\mathbf{a}}\|_2^2 \|{\mathbf{b}}_2\|_2^2 - \left| {\mathbf{b}}_2^H {\mathbf{a}} \right|^2 \right)}{\sigma^2 \|{\mathbf{a}}\|_2^2}
	\end{equation}
	\begin{equation}\label{78}
		\mathbf{F}_{\pmb{\theta}\pmb{\phi},\mathrm{FD}} = \frac{2N_a T_d\hat{p} \left( \|{\mathbf{a}}\|_2^2 {\mathbf{b}}_1^H {\mathbf{b}}_2-{\mathbf{b}}_1^H{\mathbf{a}}{\mathbf{a}}^H{\mathbf{b}}_2 \right)}{\sigma^2 \|{\mathbf{a}}\|_2^2}
	\end{equation}
	\begin{equation}\label{79}
		\mathbf{F}_{\pmb{\phi}\pmb{\theta},\mathrm{FD}} = \frac{2N_a T_d\hat{p} \left( \|{\mathbf{a}}\|_2^2 {\mathbf{b}}_2^H {\mathbf{b}}_1-{\mathbf{b}}_2^H{\mathbf{a}}{\mathbf{a}}^H{\mathbf{b}}_1 \right)}{\sigma^2 \|{\mathbf{a}}\|_2^2}
	\end{equation}
	with $\hat{p} = \frac{1}{N_a T_d} \sum_{n=1}^{N_a T_d} |\bar s(n)|^2 = \frac{1}{N_a } \sum_{n=1}^{N_a } | s(n)|^2$ being the estimate of $p = \mathbb{E} \{ |s(n)|^2 \}$. Subsequently, we can calculate $\mathbf{F}_{\pmb{\theta}\pmb{\theta}}$, $\mathbf{F}_{\pmb{\theta}\pmb{\phi}}$ and $\mathbf{F}_{\pmb{\phi}\pmb{\phi}}$ separately.
	
	For $\mathbf{F}_{\pmb{\theta}\pmb{\theta},\mathrm{FD}}$, we have $\|{\mathbf{a}}\|_2^2=N_r$, then,  $\|{\mathbf{b}}_1\|_2^2$ is computed as
	\begin{equation}\label{80}
		\begin{gathered}
			\left\|{\mathbf{b}}_1\right\|_{2}^{2}=\frac{4\pi^2\cos^2\theta\cos^2\phi}{\lambda^2}\sum_{n=1}^{N_r}{\bar{d}}_{n}^{2}
		\end{gathered}
	\end{equation}
	and  $\left| {\mathbf{b}}_1^\mathrm{H} {\mathbf{a}} \right|^2$ is:
	\begin{equation}\label{81}
		\begin{gathered}
			\left| {\mathbf{b}}_1^\mathrm{H} {\mathbf{a}} \right|^2=\frac{4\pi^2\cos^2\theta\cos^2\phi}{\lambda^2}\Bigg|\sum_{n=1}^{N_r}{\bar{d}}_{n} \Bigg|^2.
		\end{gathered}
	\end{equation}
	
Substituting   \eqref{80} and  \eqref{81}  into  \eqref{76} yields
	\begin{equation}\label{82}
		\mathbf{F}_{\pmb{\theta}\pmb{\theta},\mathrm{FD}} = \frac{8N_a T_d \hat{p}\pi^2\cos^2\theta\cos^2\phi}{N_r\sigma^2\lambda^2}\Bigg(N_r\sum_{n=1}^{N_r}{\bar{d}}_{n}^{2}-\Big(\sum_{n=1}^{N_r}{\bar{d}}_{n} \Big)^2\Bigg),
	\end{equation}
	
	Similarly, $\mathbf{F}_{\pmb{\phi}\pmb{\theta},\mathrm{FD}}$, $\mathbf{F}_{\pmb{\theta}\pmb{\phi},\mathrm{FD}}$, and $\mathbf{F}_{\pmb{\phi}\pmb{\phi},\mathrm{FD}}$ are calculated as:
	\begin{equation}\label{83}
		\hspace{-8mm}		\mathbf{F}_{\pmb{\phi}\pmb{\phi},\mathrm{FD}} = \frac{8N_a T_d\hat{p}\pi^2}{N_r\sigma^2\lambda^2}\Big(N_r\sum_{n=1}^{N_r}{\tilde{d}}_{n}^{2}-\Big(\sum_{n=1}^{N_r}{\tilde{d}}_{n} \Big)^2\Big)
	\end{equation}
	\begin{equation}\label{84}
		\hspace{-5mm}		\begin{aligned}
			&\mathbf{F}_{\pmb{\theta}\pmb{\phi},\mathrm{FD}} = \mathbf{F}_{\pmb{\phi}\pmb{\theta},\mathrm{FD}} \\
			&= \frac{8N_a T_d \hat{p}\pi^2\cos\theta\cos\phi}{N_r\sigma^2\lambda^2}\Bigg(N_r\sum_{n=1}^{N_r}{\bar{d}}_{n}{\tilde{d}}_{n}
			- \Big(\sum_{n=1}^{N_r}{\bar{d}}_{n}\Big)\Big(\sum_{n=1}^{N_r}{\tilde{d}}_{n}\Big)\Bigg),
		\end{aligned}
	\end{equation}
	
	Comparing the expression of each term in the CRLB matrix derived in Appendix A, we have
	\begin{equation}\label{85}
		{\mathop{\rm CRLB}\nolimits}_{\mathrm{HDS}}^{-1}\approx \frac{K T}{T_d} {\mathop{\rm CRLB}\nolimits}_{\mathrm{FD}}^{-1}.
	\end{equation}
	This completes the proof of Theorem 3.

\end{document}